\documentclass[sn-mathphys-num]{sn-jnl}

\usepackage{graphicx}%
\usepackage{multirow}%
\usepackage{amsmath,amssymb,amsfonts}%
\usepackage{amsthm}%
\usepackage{mathrsfs}%
\usepackage[title]{appendix}%
\usepackage{textcomp}%
\usepackage{manyfoot}%
\usepackage{booktabs}%
\usepackage{algorithm}%
\usepackage{algorithmicx}%
\usepackage{algpseudocode}%
\usepackage{listings}%
\usepackage{nomencl}
\usepackage{longtable}
\usepackage{caption}
\usepackage[table]{xcolor}
\definecolor{deepgreen}{rgb}{0, 0.4, 0} 
\definecolor{dateplum}{rgb}{0.7, 0, 0}

\usepackage{array}

\theoremstyle{thmstyleone}%
\theoremstyle{thmstyletwo}%

\theoremstyle{thmstylethree}%

\makenomenclature
\begin{document}

\title[Article Title]{Text2Scenario: Text-Driven Scenario Generation for Autonomous Driving Test}


\author[1]{\fnm{Xuan} \sur{Cai†}\footnotemark[1]}\email{caixuan@buaa.edu.cn}

\author[1,2]{\fnm{Xuesong} \sur{Bai†}\footnotemark[1]}\email{xs\_bai@buaa.edu.cn}
\footnotetext[1]{†: These author contribute equally to this work.}

\author*[1,3]{\fnm{Zhiyong} \sur{Cui}}\email{zhiyongc@buaa.edu.cn}

\author[1]{\fnm{Danmu} \sur{Xie}}\email{xiedanmu@163.com}
\author[4]{\fnm{Daocheng} \sur{Fu}}\email{fudaocheng@pjlab.org.cn}
\author[1,2]{\fnm{Haiyang} \sur{Yu}}\email{hyyu@buaa.edu.cn}
\author[1,2]{\fnm{Yilong} \sur{Ren}}\email{yilongren@buaa.edu.cn}

\affil*[1]{\orgdiv{State Key Laboratory of Intelligent Transportation Systems}, \orgname{School of Transportation Science and Technology, Beihang University}, \orgaddress{\city{Beijing}, \postcode{100191}, \state{Beijing}, \country{P.R.China}}}

\affil*[2]{\orgname{Zhongguancun Laboratory}, \orgaddress{ \city{Beijing}, \postcode{100191}, \state{Beijing}, \country{P.R.China}}}

\affil[3]{\orgname{Key Laboratory of Transport Industry of Comprehensive Transportation Theory (Nanjing Modern Multimodal Transportation Laboratory), Ministry of Transport}, \orgaddress{ \city{Nanjing}, \postcode{210000}, \state{Jiangsu Province}, \country{P.R.China}}}

\affil[4]{\orgname{Shanghai Artificial Intelligence Laboratory}, \orgaddress{ \city{Shanghai}, \postcode{200232}, \state{Shanghai}, \country{P.R.China}}}


\abstract{Autonomous driving (AD) testing constitutes a critical methodology for assessing performance benchmarks prior to product deployment. The creation of segmented scenarios within a simulated environment is acknowledged as a robust and effective strategy; however, the process of tailoring these scenarios often necessitates laborious and time-consuming manual efforts, thereby hindering the development and implementation of AD technologies. In response to this challenge, we introduce Text2Scenario, a framework that leverages a Large Language Model (LLM) to autonomously generate simulation test scenarios that closely align with user specifications, derived from their natural language inputs. Specifically, an LLM, equipped with a meticulously engineered input prompt scheme functions as a text parser for test scenario descriptions, extracting from a hierarchically organized scenario repository the components that most accurately reflect the user's preferences. Subsequently, by exploiting the precedence of scenario components, the process involves sequentially matching and linking scenario representations within a Domain Specific Language corpus, ultimately fabricating executable test scenarios. The experimental results demonstrate that such prompt engineering can meticulously extract the nuanced details of scenario elements embedded within various descriptive formats, with the majority of generated scenarios aligning closely with the user's initial expectations, allowing for the efficient and precise evaluation of diverse AD stacks void of the  labor-intensive need for manual scenario configuration. Project page: \href{https://caixxuan.github.io/Text2Scenario.GitHub.io/}{https://caixxuan.github.io/Text2Scenario.GitHub.io/}.}




\maketitle

\nomenclature{AD}{Autonomous Driving}
\nomenclature{LLM}{Large Language Model}
\nomenclature{ADS}{Autonomous Driving System}
\nomenclature{DSL}{Domain Specific Language}
\nomenclature{UML}{Unified Modeling Language}
\nomenclature{SysML}{Systems Modeling Language}
\nomenclature{SUT}{System Under Test}
\nomenclature{T2S}{Text-to-Scenario}
\nomenclature{SOTIF}{Safety of The Intended Functionality}
\nomenclature{FS}{Few-Shot}
\nomenclature{CoT}{Chain of Thought}
\nomenclature{SAC}{Syntax Alignment Checking}
\nomenclature{SC}{Self-Consistency}
\nomenclature{RQ}{Research Question}
\nomenclature{NHTSA}{National Highway Traffic Safety Administration}
\nomenclature{ADAS}{Advanced Driver Assistance System}
\nomenclature{C-NCAP}{China New Car Assessment Program}
\nomenclature{ROS}{Robot Operating System}
\printnomenclature

\section{Introduction}\label{sec1}

\subsection{Background}

The advent of AD technology marks the onset of a transformative era, promising significant advancements in traffic safety and mobility efficiency \cite{yang2019driving,peng2021uncertainty}. Prior to its widespread commercial deployment, the technology must undergo rigorous evaluations to validate its reliability and safety credentials \cite{lee2024study}. Traditional vehicle safety assessments typically depend on controlled collision experiments and behavioral testing within predefined circumstances. However, Autonomous Driving Systems (ADSs), due to their intricate operational complexities and requirements for environmental adaptability, demand the creation of a diverse array of testing scenarios to comprehensively address the inherent challenges \cite{gao2022performance}. Moreover, the vast diversity of real-world traffic scenarios suggests that physical testing alone cannot fully capture all potential risk exposures \cite{hoss2022review}. Consequently, an effective approach necessitates the utilization of computational simulation technologies to generate a broad range of intricate virtual scenarios \cite{chao2019force, wang2022parallel}. These simulated environments act as arenas for rigorously evaluating and refining the effectiveness of the ADSs. 

However, the prevailing methodologies for scripting AD test scenarios, which employ description languages such as Domain Specific Language (DSL), Unified Modeling Language (UML), and Systems Modeling Language (SysML), remain largely manual and require continuous maintenance \cite{meyer2021simulator}. This traditional approach presents considerable deficiencies and limitations. To begin with, the manual composition of DSLs for intricate, multi-actor scenarios constitutes a laborious and resource-intensive tasks, with developers committing substantial effort to guarantee scenario accuracy, consistency, and comprehensiveness. Moreover, manually crafted DSLs are prone to errors and oversights, potentially leading to insufficient test coverage or a reduction in testing rigor. For instance, scripting a basic car-following scenario may involve more than 200 lines of DSL code \cite{lou2022testing}, requiring considerable time for understanding and maintenance. This process is particularly challenging for novices, presenting a steep learning curve. 

An additional critical concern is the rapid advancement of AD technology, which simultaneously intensifies the demand for a continuously evolving set of test scenarios. The process of manually maintaining and updating DSLs to keep pace with these advancements is formidable and fraught with challenges, often resulting in the obsolescence and functional deficiencies of DSLs. Moreover, reliance on individual expertise and domain-specific insight during the DSL authoring process results in significant variability across scenarios crafted by different developers, thereby compromising standardization and reusability \cite{li2022domain}. As a result, the industry urgently requires an automated DSL generation mechanism —- one that comprehends developers' intentions while offering convenience, automation, standardization, and reusability. 

In response to these challenges, scholarly endeavors have initiated the exploration of methodologies for the automatized generation of DSLs tailored to AD testing scenarios. LawBreaker \cite{sun2022lawbreaker} innovatively converts traffic regulations into driver-centric Signal Temporal Logic specifications and employs a fuzzy testing engine to uncover diverse modalities of rule violation by optimizing specification coverage. However, the initialization of the scenarios persists as a manual process. Contrarily, the pioneering world model, DriveDreamer-2 \cite{zhao2024drivedreamer} circumvents the need for conventional DSL representations and leverages LLMs to decipher user requirements, generating detailed multi-perspective video scenarios via diffusion models. Notwithstanding, this approach does not facilitate a closed-loop interaction between the tested AD vehicle and the traffic environment, and its application is confined to the utilization in end-to-end ADS, with planning as the central objective \cite{hu2023planning}. 

\subsection{Research Process}
To reconcile the automation imperatives with empirical research advancements, we introduce an innovative framework for automated simulation testing scenario generation,  referred to as Text2Scenario (T2S). Initially, we deconstruct complex traffic scenarios into fundamental components and establish a hierarchical scenario repository, ensuring complete coverage of scenario elements for the construction of the material library. Subsequent to that, we meticulously engineered a prompt-driven workflow and devised a testing text parser predicated on an LLM, improving its ability to comprehend and interpret natural language descriptions of test scenarios. Specifically, to navigate the intricacies and ambiguities inherent in natural language, the LLM employs a multi-stage in-context few-shot learning approach. This methodology allows for the direct selection of congruent components from the scenario material repository that align with the textual descriptions, thus enabling the seamless generation of accurate scenario representations. 

Following this, we construct a customized DSL corpus -- encompassing controllable scenario parameters and events -- to function as a reliable material repository for the targeted scenario description document. Grounded in the scenario representation, we leverage static element matching and dynamic element concatenation techniques within a priority-based assembly architecture. This approach carefully modulates the scenario parameter values within the seed DSL document and incorporates segments for event management, leading to the creation of a precise and standardized scenario description file. To conclude the process, we tailor evaluation indicators specific to the ADS under test. These metrics facilitate the real-time evaluation of the ADS within the simulation platform, ultimately yielding comprehensive testing reports. 

\subsection{Contribution}
Our contribution lies in the three folds: 

\begin{itemize}
\item {We introduce a novel framework, Text2Scenario (T2S), for automated testing scenario generation, which is segmented into five distinctive stages and predicated on textual descriptions to enable virtual simulation testing of ADS;}

\item {To the best of our knowledge, our work represents the pioneering effort in taking advantage of the capabilities of LLM for parsing intricate natural language scenario descriptions within the realm of DSL and the subsequent generation of standardized, manipulable scenario representations;}

\item {Harnessing a spectrum of test scenario descriptions, we conducted an extensive series of simulation experiments on the Carla platform, which led to the identification of 533 driving safety violations in ADS across 368 generated simulation scenarios. A comprehensive analysis of these test reports is instrumental in revealing technical vulnerabilities in ADS.}
\end{itemize}

The structure of this paper is organized as follows. A critical review of the pertinent literature is conducted in Section 2. Section 3 introduces our proposed framework for generating test scenarios. The research questions and architecture of the experimental setup come under scrutiny in Section 4. We share the results of our experimental investigations in Section 5, followed by a discussion on threats to validity of the work in Section 6. In the concluding Section 7, we synthesize the findings of the study and propose directions for future works.

\subsection{Motivation}\label{sec2}
The current landscape of AD scenario testing is encumbered by labor-intensive and time-consuming manual processes that involve translating functional scenarios detailed in natural language into explicit specific test parameters \cite{menzel2018scenarios,zhang2022performance}, encoded into error-free DSL files \cite{asam_openscenario_2021}. There is an imperative requirement for an automated and controllable simulation scenario generation methodology that enhances scalability, open-endedness, standardization, and compatibility. 

On the contrary, LLMs with powerful natural language generation capabilities \cite{xu2022systematic}, offer an automated solution, capable of rapidly generating complex and varied test scenario representations based on succinct text prompts. When juxtaposed with traditional manual and simulation methods, LLMs wield significant advantages such as: 

\begin{itemize}
\item {{\textit{Efficiency:}} the ability to swiftly produce a substantial volume of test scenarios;}

\item {{\textit{Diversity:}} the generation of a broad spectrum of scenarios ensuring extensive coverage;}

\item {{\textit{Interpretability:}} formulation in natural language that is straightforward to comprehend and modify;}

\item {{\textit{Cost-effectiveness:}} obviating the need for extensive manual efforts.}
\end{itemize}

In summary, LLMs hold considerable promise in supplanting the manual interpretation of human natural language, fostering standardized and adjustable scenario representations, thereby substantially expediting the testing processes for ADS.

\section{Literature Review}\label{sec6}
\subsection{Scenario-based ADS Testing}\label{sec6.1}

Autonomous vehicles' development and deployment have intensified the need for robust validation and verification methods \cite{pathrudkar2023safr}. Among the various approaches proposed in the literature \cite{aparow2021scenario,koopman2017challenges,indaheng2021scenario}, one particularly promising avenue is \textbf{scenario-based testing}. By formulating scenarios that encompass diverse real-world situations, researchers aim to push the boundaries of testing methodologies. 

One of the key aspects of scenario-based testing is the generation of relevant test scenarios. Ghodsi et al. \cite{ghodsi2021generating} propose an efficient mechanism to characterize and generate testing scenarios using a state-of-the-art driving simulator. Wang et al. \cite{wang2022autonomous} utilize in-depth crash data involving powered two-wheelers to generate realistic testing scenarios for ADSs. Several other works have focused on the generation of challenging scenarios for AV testing. Zhou et al. \cite{zhou2022genetic} employ a genetic algorithm to generate scenarios that increase the probability of collisions or near-misses, which are deemed as challenging for AVs. Chen et al. \cite{chen2021adversarial} introduce an adversarial evaluation framework that generates lane-change scenarios to expose weaknesses in AVs' decision-making policies.

In addition to scenario generation, researchers have explored methods for efficiently searching the vast scenario space to identify critical test cases. Feng et al. \cite{feng2022multimodal} propose a multimodal critical-scenario search method that combines optimization techniques with supervised learning to efficiently find scenarios that expose AV failures. In order to expand the security testing theory in connected environments, Shi et al. \cite{shi2022integrated} present an integrated traffic and vehicle co-simulation framework that enables testing of connected and autonomous vehicle technologies, including vehicle-to-vehicle and vehicle-to-infrastructure communication. 

Current research predominantly revolves around the identification and analysis of risk scenarios within AD through refinement and application of optimization or learning algorithms, leveraging existing datasets. Yet, this conventional approach necessitates user intervention for the initial establishment of seed scenarios, resulting in a notable absence of fully automated processes capable of producing anticipated scenario configurations derived directly from user-conceived ideas or blueprints.

\subsection{LLM-Driven Scenario Generation}\label{sec6.2}

The advent of advanced large model technologies in recent years has indeed unlocked new possibilities in automatically generating high-quality datasets, inclusive of input scenarios pertinent to AD \cite{xi2023rise}. An emerging research endeavor involves employing LLMs for the simulation and generation of traffic scenarios with increased fidelity. The CTG++ \cite{zhong2023language} framework devised by Zhong et al. harnesses the synergy between spatiotemporal transformers and LLMs, empowering users to craft realistic and controllable traffic scenarios via intuitive natural language instructions. Moreover, Li et al. \cite{li2024scenarionet} have showcased the potential of LLMs in conjuring traffic scenarios from SUMO configuration files, effectively circumventing the conventional reliance on graphical editor interfaces or the labor-intensive process of manual XML file authorship. Despite these advancements, the extent of controllability within these generated scenarios remains an area deserved for further exploration and validation.

Recent research contributions reveal an intensified interest in harnessing LLMs' capabilities to assimilate traffic regulations and natural language descriptions, transforming them into explicit driving scenarios. Deng et al. \cite{deng2023target} and Guezay et al. \cite{guzay2023generative} pioneered methods wherein LLMs facilitate the automatic distillation of knowledge from traffic laws, spawning driving scenarios congruent with regulatory mandates, which subsequently serve in assessing diverse ADS software stacks. This innovation significantly bolsters automation and broadens the variety of AD test scenarios \cite{lykov2023llm}. In a similar vein, Barone et al. \cite{miceli2023dialogue} ventured into merging LLMs with automated driving scenario generation, elucidating techniques to construct and refine the driving conduct of AI-powered vehicles predicated on natural language scripts. Extending the application of LLMs beyond the automotive sphere, Cao et al. \cite{cao2023robot} proposed a novel approach to craft robot behavior trees. Their methodology leverages LLMs for designing and autonomously realizing intricate cross-domain tasks through an intelligible behavior tree architecture. 

While the domain of scenario generation leveraging LLMs is experiencing rapid growth, it currently faces a shortfall in tailored research aimed at crafting scenarios that align squarely with user expectations and benefit from LLM-assisted unified DSL creation. Distinct from prevailing studies in the arena of scenario generation, our approach capitalizes on the capabilities of LLMs to produce standardized, manageable, and universally applicable scenario description files via linguistic and textual inputs. This methodology ensures compatibility across a broad spectrum of test subjects and environments, marking a significant advancement towards more versatile and user-centric scenario generation frameworks.

\section{Methodology}\label{sec3}

This section delineates the proposed framework, T2S, for testing scenario generation. Commencing with the insertion of the testing language text into the LLM-empowered parser, it comprehends the user's specifications and outputs the scenario representation leveraging the hierarchical scenario repository. Following this, the DSL-based scenario generator is employed to fabricate standardized scenario files. These files are subsequently fed into the simulation environment, whereupon critical evaluative metrics are real-time monitored. 

\subsection{Overview}\label{subsec3.1}

\begin{figure}[t!]
\centering
\includegraphics[width=13cm]{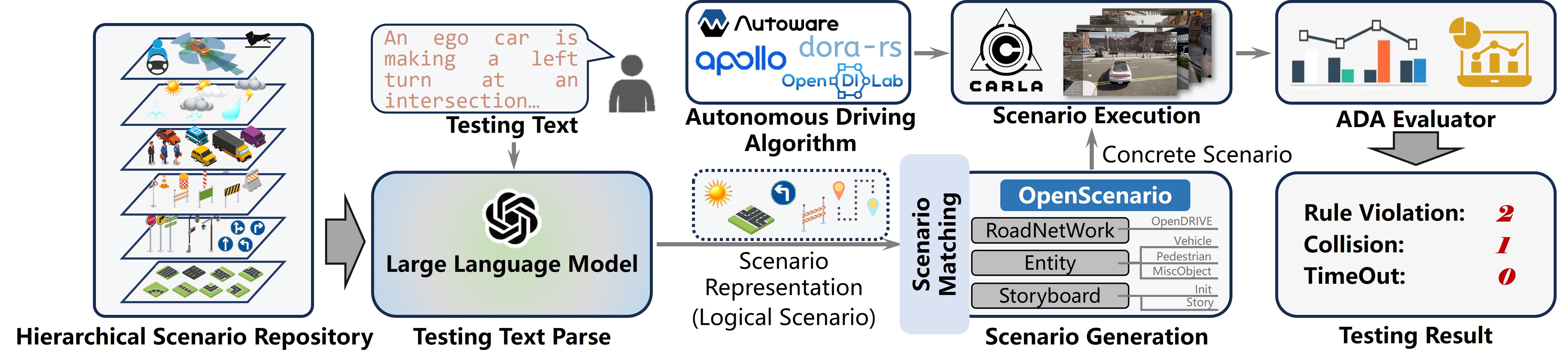}
\caption{Overview of the T2S testing scenario generation framework.}
\label{fig_overview}
\vspace{-0cm}
\end{figure}

Fig.\ref{fig_overview} elucidates the T2S testing scenario generation framework advanced in this research. The T2S system transmutes testing descriptive text into simulator-ready DSL scripts across five distinct stages, anchored by the ASAM OpenScenario \cite{asam_openscenario_2021} textbook approach. The first stage necessitates listing the sextet of pivotal traffic scenario elements, in accordance with the Safety of The Intended Functionality (SOTIF)\cite{wu2021new} tenets, which will be exhaustively expounded upon in Section \ref{subsec3.2}. Notably, vehicle-to-everything communication level is exempted from this discourse, as it falls outside the scope of this paper.

The secondary stage involves the LLM decoding user-provided testing text into logical scenarios by the knowledge base of scenario descriptions, identifying the relevant hierarchical pre-configured elements. The following stage proceeds to match the scenario primitive, employing a priority-based assembling methodology to concatenate scenario descriptors into a coherent DSL script, \textcolor{black}{thus converting these logical scenarios into concrete, executable scenarios.}

Ultimately, this scenario descriptor file (DSL) is integrated into the simulation platform, which interprets and executes the scenario. This integration supports the validation and verification of various AD stacks during the execution phase to yield testing report results. 

\subsection{Hierarchical Scenario Repository}\label{subsec3.2}

Given numerous intricate components within traffic scenarios, such as ambient weather, road topologies, and unforeseen road debris, it becomes imperative to systematically categorize these elements to define and confine them to a controllable domain concisely \cite{wang2022autonomous}. Nonetheless, the quantification and description of many such elements present significant challenges \cite{wen2020scenario}, particularly regarding the subjectivity of participants' yielding behaviors or the unpredictable driving styles of individual drivers. Conventional DSLs \cite{fremont2019scenic, queiroz2019geoscenario} necessitate precise and stringent parameter configurations to construct fully functional logical or concrete scenarios. Such rigidity in data specification inadvertently restricts the scenario's malleability, consequentially limiting the exploration of potential flaws in the system under test (SUT). To surmount this limitation, we advocate for a more versatile scenario description language designed to enhance adaptability, empower exploratory changes, and potentially transcend the pre-established confines of stringent DSL scripts at the nascent stages.

\begin{figure}[t!]
\centering
\includegraphics[width=5cm]{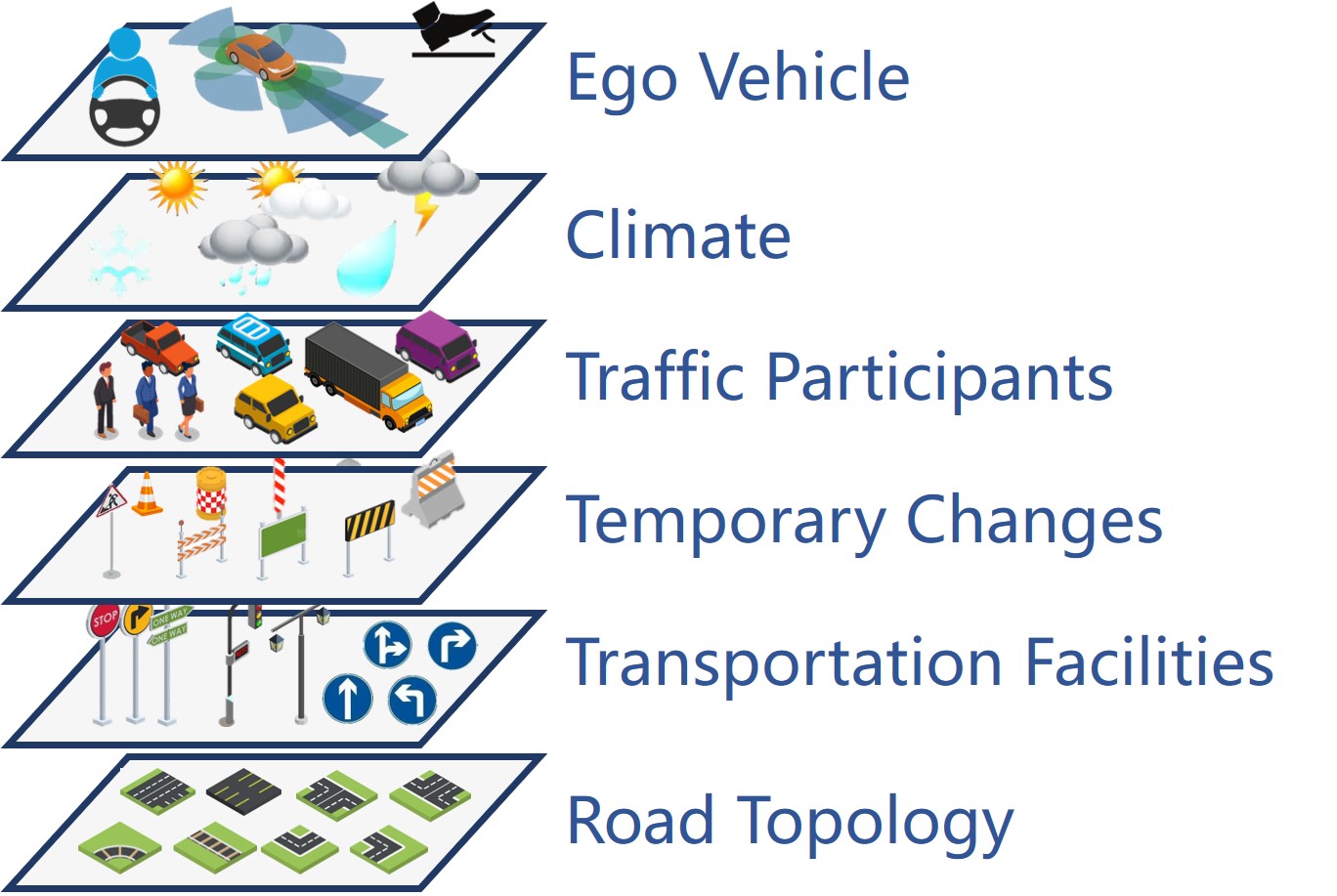}
\caption{Hierarchical scenario description. The communication status level is ignored from consideration as our focus is exclusively on the self-driving AD stack.}
\label{fig_hierarchical}
 \vspace{-0cm}
\end{figure}

The SOTIF-based scenario conceptualizes a hierarchical scenario framework in a systematized and canonical format, delineating interconnections between diverse components, thereby engendering logical affiliations and a tiered structure. Integration of static and dynamic elements is imperative, as the latter is contingent upon the former for contextual grounding. The communication status is intentionally discounted, given our singular emphasis on the self-driving entity. The 6-tiered edifice of hierarchical scenario depiction is assembled layer upon layer, ascending from foundational components to nuanced particulars, as depicted in Fig.\ref{fig_hierarchical}. Leaning on the pre-crash documentation by the National Highway Traffic Safety Administration (NHTSA) \cite{najm_pre-crash_2007}, OpenXOntology \cite{asam_openxontology_2021}, and the Traffic Safety Handbook \cite{texas_dmv_handbook_2022}, Tab.\ref{tab_element} enumerates potential elements for every component -- road topology, for instance, might encompass intersections, roundabouts, and T-junctions. These elements harbor the capacity to encapsulate extensive semantic connotations, embracing the majority of typical scenario constituents encountered in the real world. Their ordered combination yields a comprehensive emulation of virtually any envisaged target scenario.

\begin{table}[!t]
\centering
\caption{Examples of optional elements for hierarchical scenario repository.}
\label{tab_element}
\begin{tabular}{p{0.2\linewidth}|p{0.7\linewidth}}
\hline
\textbf{Component} & \textbf{Element} \\ \hline
Road Topology & \begin{tabular}[t]{@{}l@{}}
Topology: intersection, roundabout, T-junction, etc.\\
Lanes: single lane, two-lanes, three-lanes, etc.
\end{tabular} \\ \hline
Transportation Facilities & \begin{tabular}[t]{@{}l@{}}
Road Markers: solid line, double solid line, broken line, etc.\\
Traffic Signs: traffic light, stop sign, speed limit sign, etc.
\end{tabular} \\ \hline
Temporary Changes & \begin{tabular}[t]{@{}l@{}}
Type: cone barrel, warning sign, warning bucket, etc.\\
Position: front, left, right, etc.
\end{tabular} \\ \hline
Traffic Participants & \begin{tabular}[t]{@{}l@{}}
Type: car, truck, van, etc.\\
Position: front, left, right, etc.\\
Oracle: longitudinal (yield, accelerate, decelerate), lateral (keep lane, \\change lane), global behavior (go forward, turn)
\end{tabular} \\ \hline
Climate & \begin{tabular}[t]{@{}l@{}}
Weather: type (sunny, rainy, snowy) and density (strong, medium, weak)\\
Time: daytime, nighttime, morning, etc.
\end{tabular} \\ \hline
Ego Vehicle & \begin{tabular}[t]{@{}l@{}}
Type: car, truck, van, etc.\\
Position: roadside, right lane, left lane, etc.\\
Global Behavior: go forward, turn left, turn right, etc.
\end{tabular} \\ \hline
\end{tabular}
\end{table}

\begin{itemize}
\item {{\textit{Road Topology:}} Establishes the foundation upon which all scenario elements rest;}

\item {{\textit{Transportation Facilities:}} Erected atop road topologies, these structures facilitate transport;} 

\item {{\textit{Temporary Changes:}} Temporary modifications to road structures and transportation facilities; }

\item {{\textit{Traffic Participants:}} Encompasses all entities within the background traffic flow, characterized by their static attributes and dynamic behaviors;}

\item {{\textit{Climate:}} Represents the array of environmental factors within the scenario that influence the preceding components as well as the functionality of vehicle entities; }

\item {{\textit{Ego Vehicle:}} Leverages the constructed external scenario's first five tiers to influence information and actions occurring within the vehicle's interior; }

\end{itemize}

\subsection{Testing Text Parsing}\label{subsec3.3}

Building on unparalleled advancements in context comprehension, textual synthesis, and image-text correspondence, LLMs like GPT-4 \cite{openai_chatgpt_2022} have ascended at a trajectory potentially rivaling or surpassing human expertise. LLMs are on the cusp of supplanting human roles in knowledge interpretation, contextual alignment, and generation of experiential narratives \cite{laskar2023systematic} within the domain of standardized process creation. Consequently, within the T2S framework, we investigate the deployment of LLMs as testing text parsers, transfiguring user-generated linguistic descriptions into definitive scenario representations. 

\begin{figure}[t!]
\centering
\includegraphics[width=13cm]{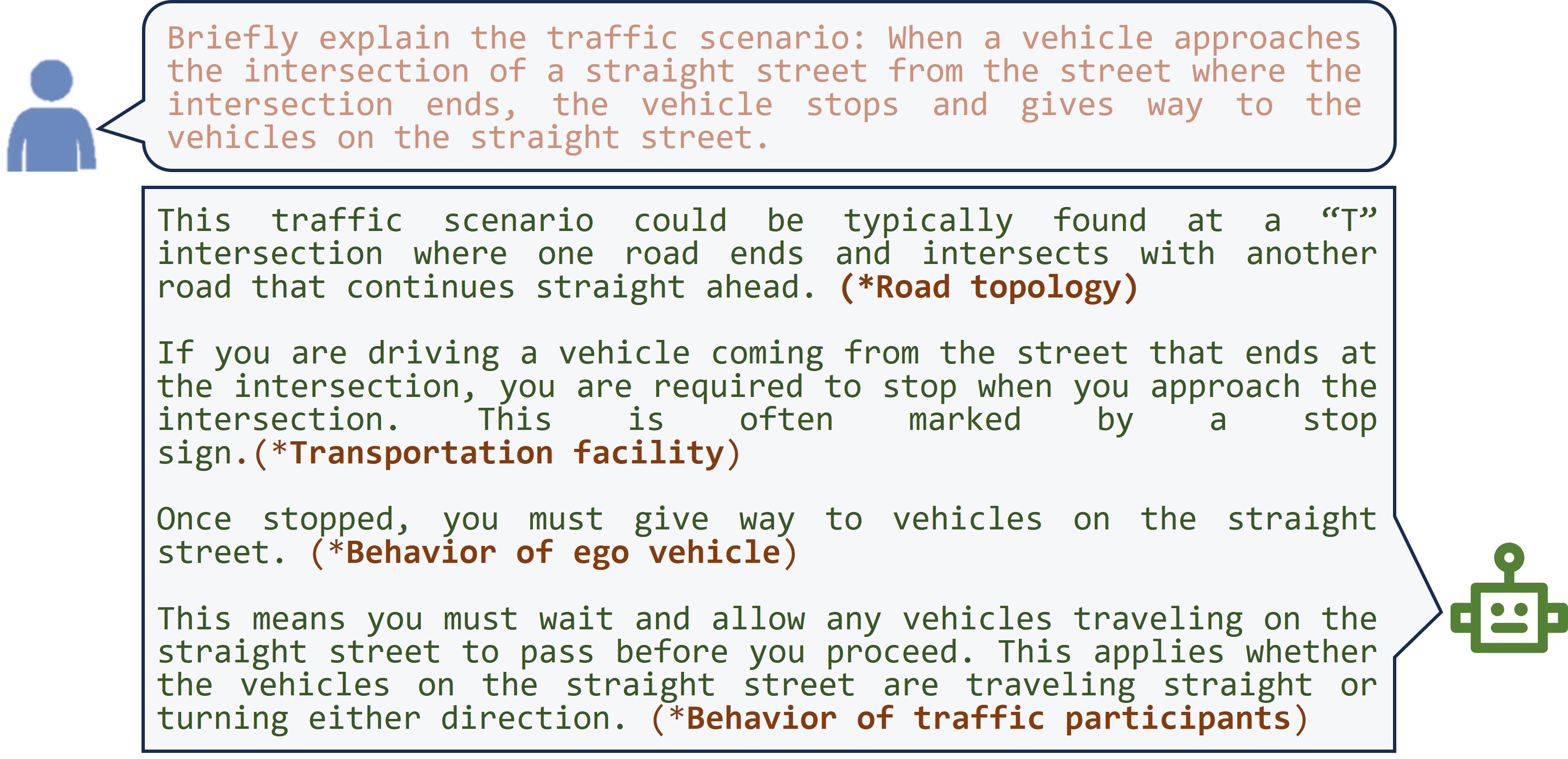}
\caption{A case of using prompt for interaction with GPT-4.}
\label{fig_case}
\vspace{-0cm}
\end{figure}

(1) \textit{Prompt engineering:} Following thorough training, LLMs have amassed a knowledge repository approximating human expertise. A minimal prompt can actuate these LLMs to assimilate new information and exhibit user-anticipated outputs without necessitating alterations to their intrinsic weighting. This necessitates meticulous prompt engineering to remold the functionality of LLMs. 
In one case study, we illustrate the interaction with GPT-4 by inputting a scenario description prompt. Referenced in Fig.\ref{fig_case}, the case study demonstrates that GPT-4 is capable of apprehending a succinct scenario narrative and identifying critical elements within. GPT-4's ability to infer underlying road topologies, such as a "T" intersection, and discern the dynamics of transportation facilities, ego vehicle behavior, and traffic participant activities from a concise description is based on prior-knowledge.

Yet, while LLMs can distill key information from natural language, their linguistic outputs are not currently suitable for direct use in constructing testing scenarios. The incomplete coverage of components in Tab.\ref{tab_element} and the introduction of uncertain elements may prevent executable scenarios or produce imprecise guidance, leading to distortions. Furthermore, the conversion of unbounded unstructured output into the hard code format is challenging. 

\begin{figure}[t!]
\centering
\includegraphics[width=13cm]{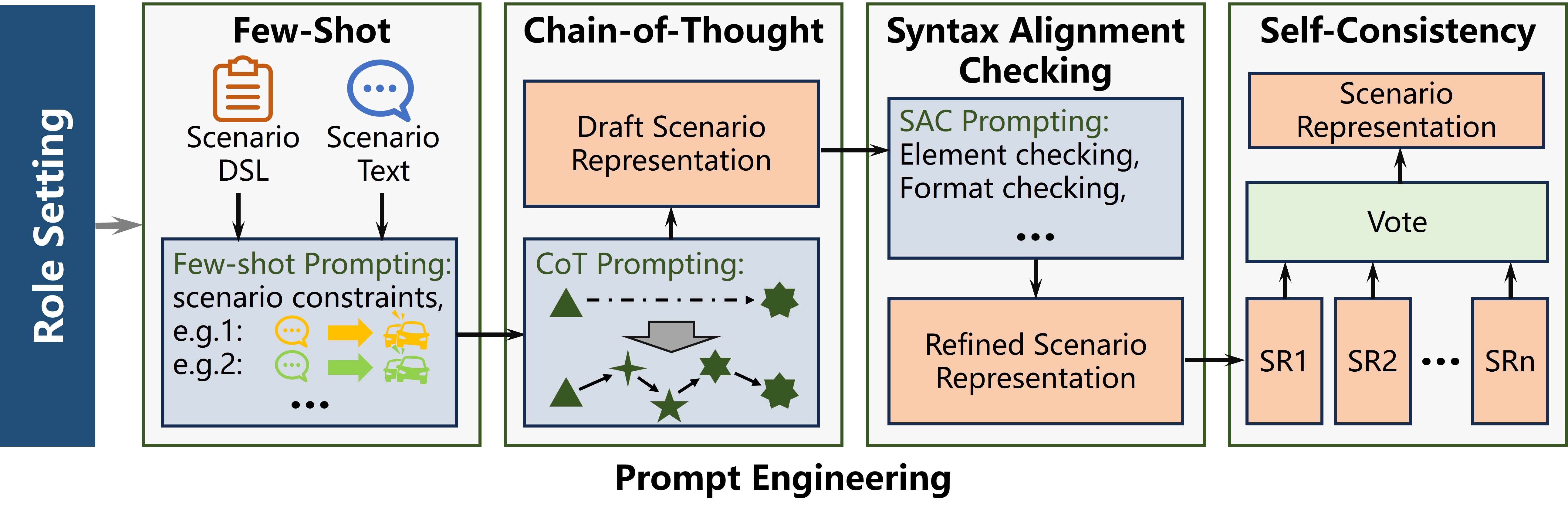}
\caption{Pipeline of the prompt engineering embedded in testing text parsing. The Prompt Engineering process encompasses five stages: Role Setting, Few-Shot, Chain-of-Thought, Syntax Alignment Checking, and Self-Consistency. The text description undergoes sequential prompt processing to yield a scenario representation.}
\label{fig_engineering}
\vspace{-0cm}
\end{figure}

\begin{figure}[t!]
\centering
\includegraphics[width=13cm]{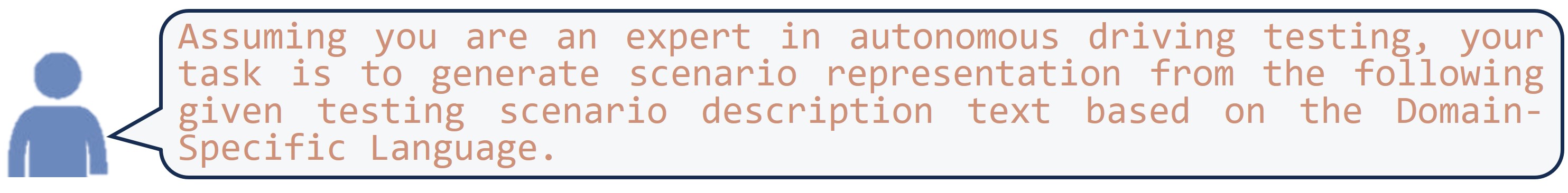}
\caption{Prompt of role setting of LLM.}
\label{fig_role}
\vspace{-0cm}
\end{figure}

To address these issues, we refined a prompt pipeline tailored for testing text parsers, depicted in Fig.\ref{fig_engineering}. This pipeline guides LLMs in generating structured outputs that align with DSL requirements. The Prompt Engineering process encompasses five stages: Role Setting, Few-Shot, Chain-of-Thought, Syntax Alignment Checking, and Self-Consistency. Each stage employs carefully engineered prompts to elicit stable output from the LLM. Within the first stage, Role Setting acts as the LLM's prefix prompt, positioning the LLM as an AD testing expert to refine knowledge retrieval and enhance the quality of responses, an example of which is shown in Fig.\ref{fig_role}.

\begin{figure}[t!]
\centering
\includegraphics[width=13cm]{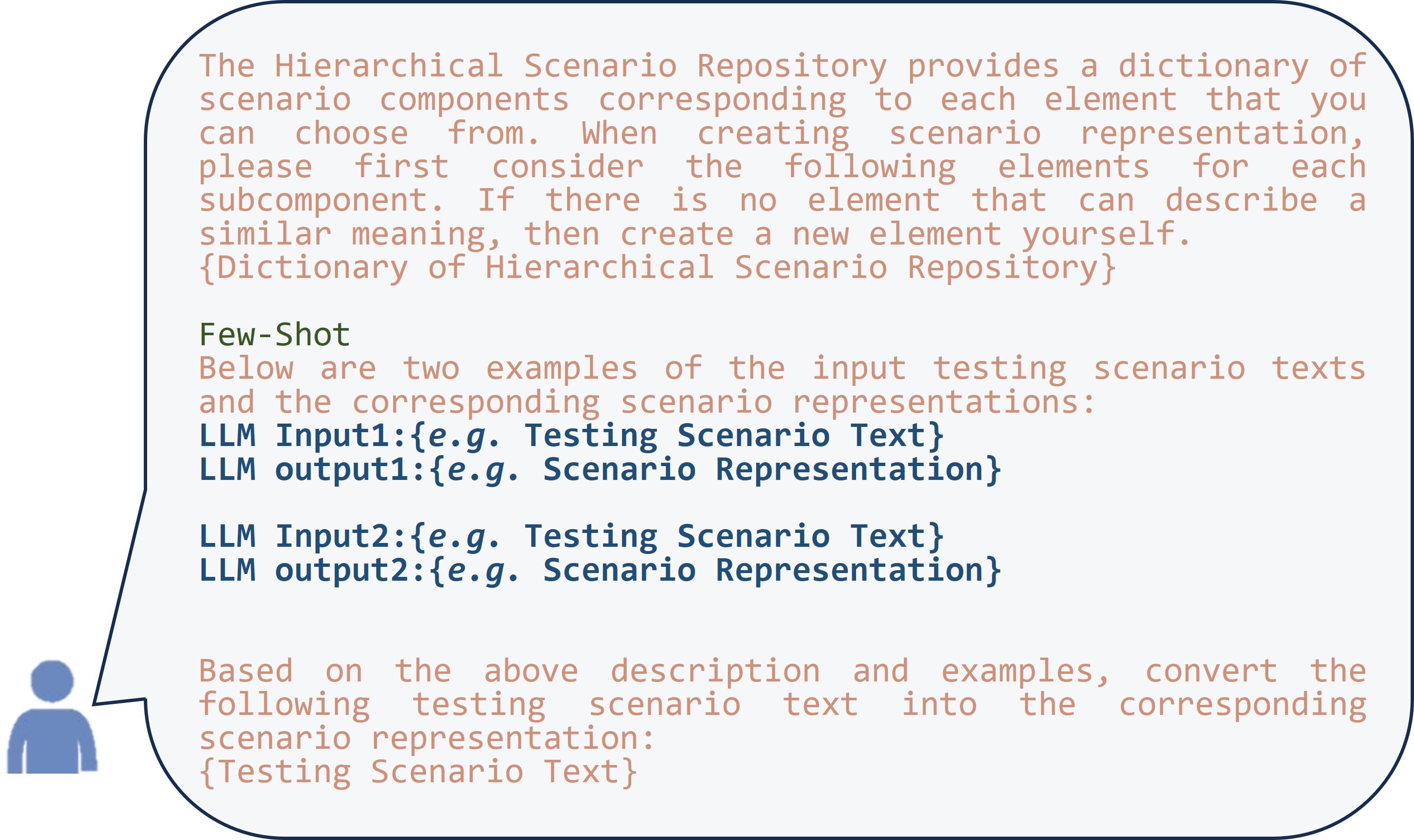}
\caption{Prompt of few-shot of LLM. The primary objective of few-shot learning is to furnish LLM with examples of inputs and outputs beforehand, aiding it in discerning the intrinsic patterns between inputs and outputs.}
\label{fig_few_shot}
\vspace{-0cm}
\end{figure}

(2) \textit{Few-Shot (FS):} To construct the foundational prompt that harnesses in-context few-shot learning paradigm \cite{liu2024visual}, we initiate by embedding the scenario repository as the selectable knowledge base for scenario elements within the LLM. This approach enables the LLM to comprehend and reproduce structured scenario representations. We facilitate this learning process by supplying a series of input-output case pairs, each consisting of scenario narrative text and its corresponding structured representation. These example pairs act as templates, providing the model with paradigms of standardized structured outputs. Ultimately, we introduce the scenario narrative that necessitates translation into a structured format. This text serves as the basic prompt trigger, which, when combined with the aforementioned input-output examples, is furnished to the LLM. This process is graphically encapsulated in Fig.\ref{fig_few_shot}, illustrating how the LLM synthesizes the provided information to generate structured scenario representations.

(3) \textit{Chain-of-Thought (CoT):} Robust logical reasoning stands as a pillar of the “Intelligence Emergence" exhibited by LLMs \cite{wei2022chain}, with the crux of reasoning hinging upon the enhancement of their thought processes. Reliance solely on an LLM's inherent knowledge repository may prove inadequate for resolving emergent and unique challenges. Frequently, the diminished certainty in an LLM's outputs can be attributed to intricate cognitive operations challenging to navigate for sophisticated tasks. For instance, when tasked with mapping the scenario "unprotected left turn for traffic vehicle" into a corresponding scenario representation, LLMs might not autonomously infer the human-favored action "yield". Instead, it may default to a "decelerate" action, erroneously conflating lower velocities as yielding. 

Nonetheless, the capability of LLMs to assimilate new knowledge through systematic prompt induction -- without alterations to their internal weights -- is significant. By instilling fundamental thinking steps akin to those of a novice, LLMs can be trained to produce outputs that more closely align with human reasoning, thus unraveling complex tasks beyond the scope of mere few-shot prompt training. 

\begin{figure}[t!]
\centering
\includegraphics[width=13cm]{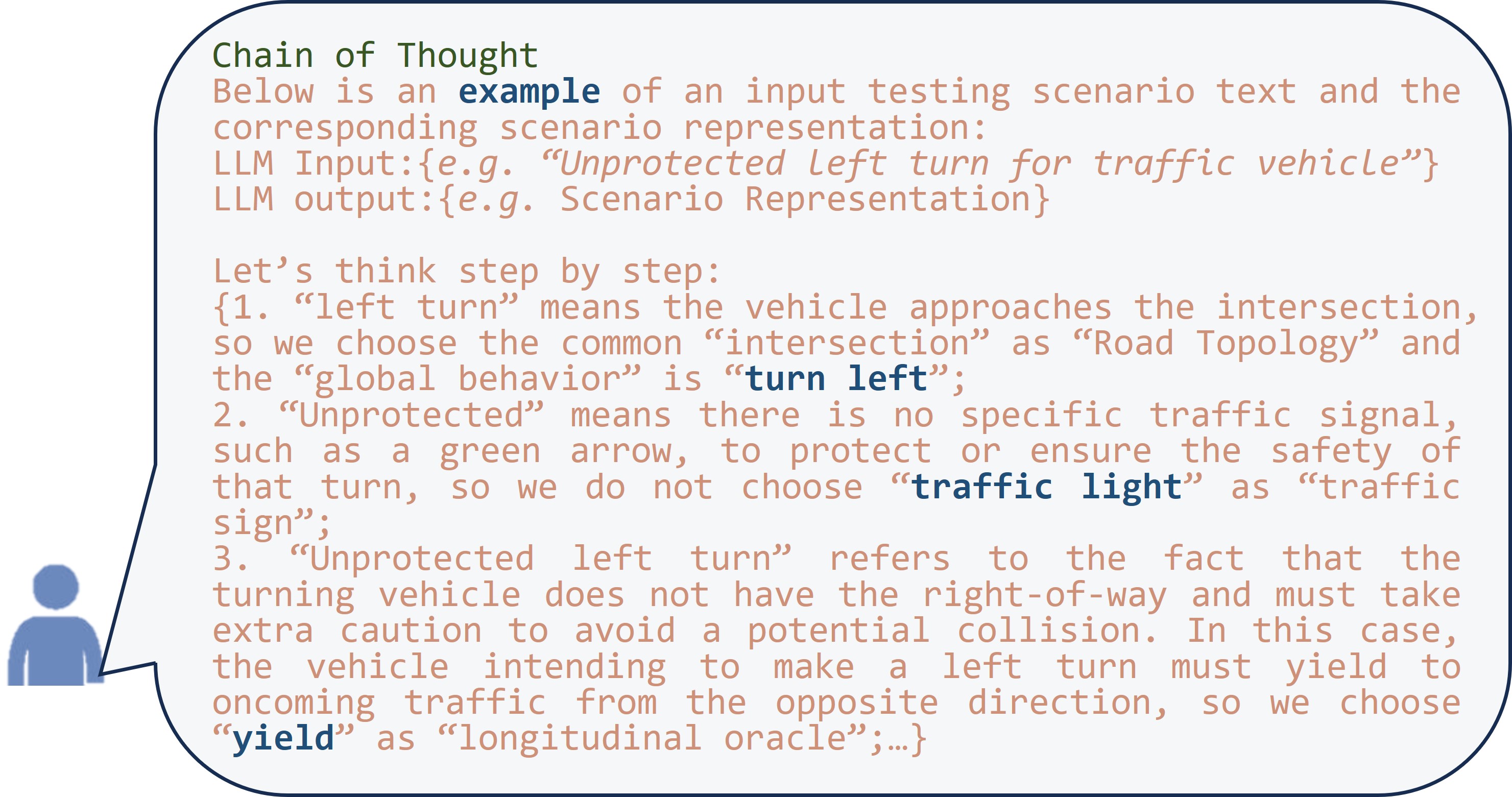}
\caption{Prompt of chain-of-thought of LLM. The core mission of CoT is to present a case study of input and output, subsequently directing LLM to emulate the process of deconstructing complex tasks by mirroring human thought processes.}
\label{fig_chain_of_thought}
\vspace{-0cm}
\end{figure}

As illustrated in Fig.\ref{fig_chain_of_thought}, we dissect the given scenario text "Unprotected left turn for traffic vehicle," extracting pertinent terms and individually scrutinizing them to aid the LLM in a more deliberate contemplation of the natural language stimuli presented. This careful analysis lays the groundwork for the LLM to formulate an appropriate scenario representation, hewing closely to the specified testing scenario via tailored instructional prompts.

(4) \textit{Syntax Alignment Checking (SAC):} Upon the generation of a scenario representation by the LLM, it is imperative to conduct both knowledge validation and syntax harmonization on the output. Knowledge validation involves cross-examining the LLM's conceptual grasp against the original scenario text to ensure congruity. Concurrently, syntax harmonization is employed to refine the output into structured data. To illustrate, an LLM may render the understanding of "straight forward" in a given scenario, while the equivalent term within the scenario repository might be cataloged as "go forward". This minor discrepancy, as minute as a singular word, could thwart hard-coded matching algorithms from accurately identifying the corresponding DSL corpus.

While precision in matching terms is crucial, it is also essential to preserve the intrinsic adaptability of the LLM. The model's unique cognitive potential may extend beyond human contemplation, permitting insight into innovative traffic elements previously unconsidered. In instances where the LLM upholds its original output due to the absence of comparable semantics within the repository, such uniqueness should be evaluated, not immediately overridden. Furthermore, the structuring of LLM output into a definitive data format is vital, sidestepping the need for laborious manual alignment during intermediary stages. It ensures that the outputs are not only accurate and reflective of the LLM's intelligence but also readily assimilable within the targeted simulation frameworks.

\begin{figure}[t!]
\centering
\includegraphics[width=13cm]{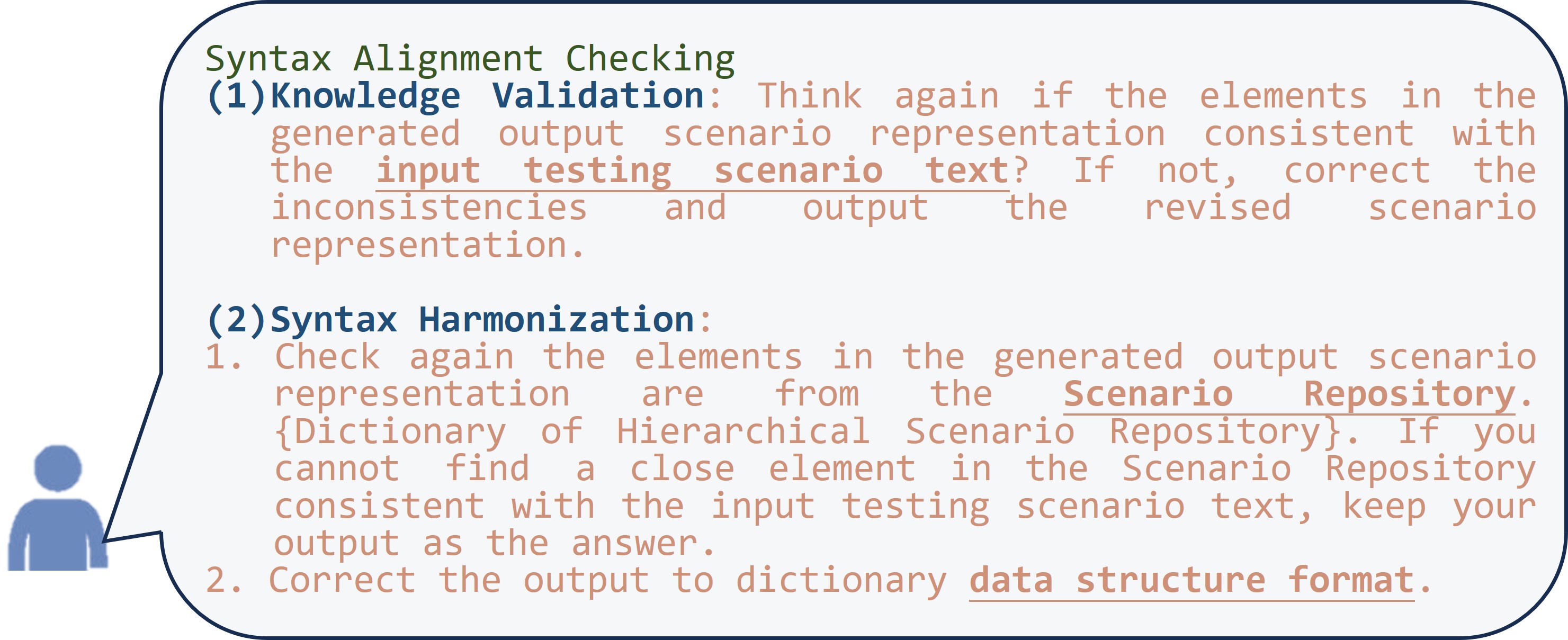}
\caption{Prompt of syntax alignment checking of LLM. SAC encompasses components knowledge validation and syntax harmonization, which are tasked with ensuring semantic consistency, representation matching and structuring, respectively.}
\label{fig_syntax_alignment_checking}
\vspace{-0cm}
\end{figure}

\begin{figure}[t!]
\centering
\includegraphics[width=13cm]{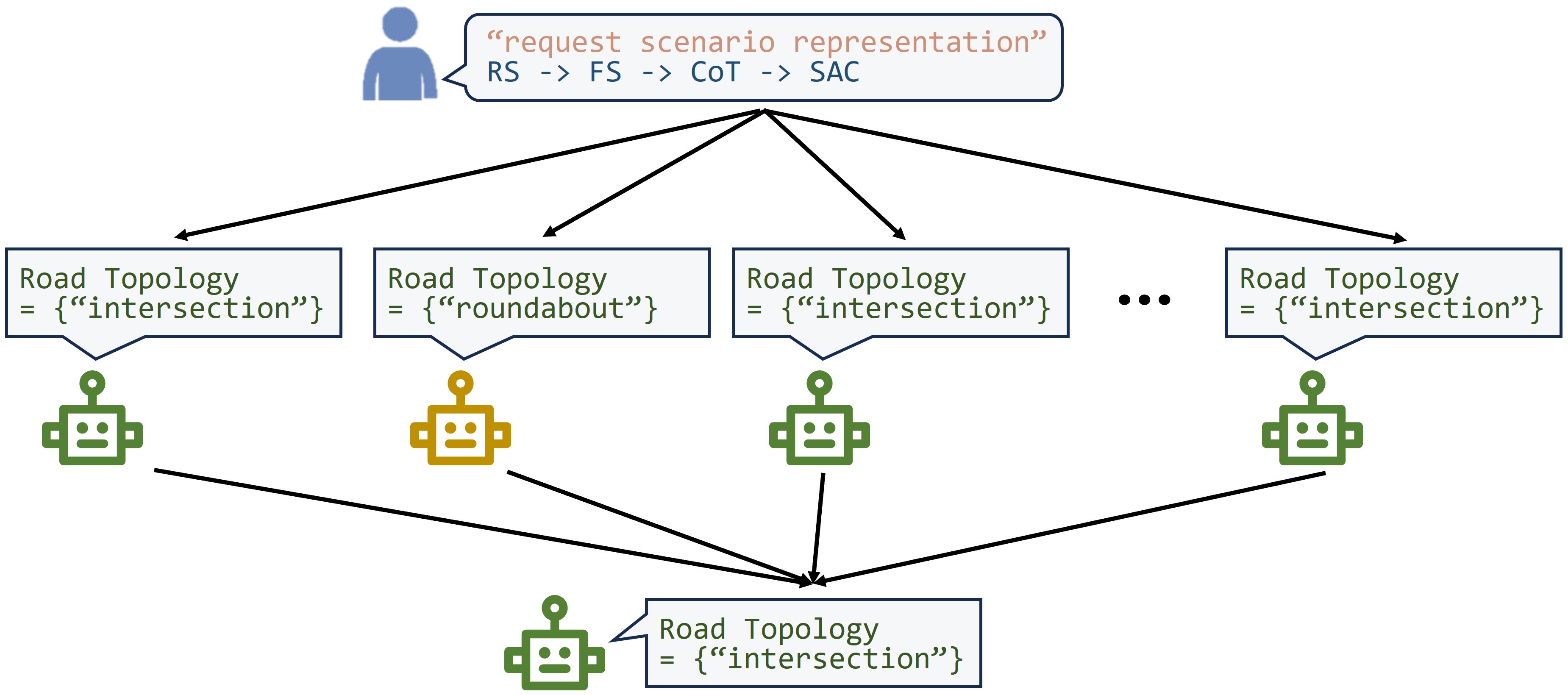}
\caption{A case of self-consistency of LLM. Diverse inference chains can result in varied representation outputs due to the randomness. The objective of self-consistency involves selecting the majority one through a voting mechanism. \textcolor{black}{Ten is chosen as the number of inference paths \cite{wang2022self}}.}
\label{fig_self-consistency}
\vspace{-0cm}
\end{figure}

(5) \textit{Self-Consistency (SC):} Given the inherently stochastic nature of deep learning models, the outputs produced by an LLM may exhibit a degree of randomness, leading to potential misalignments with human performance expectations. This entails a balance between response diversity and stability, modulated by hyperparameters such as temperature settings and top-k (nucleus sampling) criteria. To tackle this variability, a self-consistency strategy is employed. This approach involves the generation of several independent reasoning chains from the LLM. Diverse responses are extracted from each chain and the predominant result is determined through a consensus mechanism (or rather, majority vote \cite{wang2022self}), drawing parallels with the deep ensemble technique \cite{rahaman2021uncertainty} in deep learning to mitigate the unpredictability inherent in a singular model output. The concrete steps consist of drawing distinct inference trajectories from the LLM, followed by a marginalization process that consolidates responses to yield a final answer. As depicted in Fig.\ref{fig_self-consistency}, the graphical illustration represents the self-consistency approach, where multiple divergent reasoning pathways converge to a dominant output labeled as the "intersection". By canvassing a spectrum of potential solutions, the LLM increases the likelihood of generating accurate or significant responses to eliminate potential biases, particularly when confronted with complex tasks that challenge CoT techniques.

\subsection{Domain-Specific Language-based Scenario Generation}\label{subsec3.4}

Acquisition of a DSL-encoded scenario description file amenable to execution by AD simulation software necessitates a systematic translation of the LLM-output scenario representation. This paper opts for the \textit{.xosc} file format, adhering to the ASAM OpenScenario standard. The task involves meticulous link-tuning, effectively correlating the key-value pairs present within the scenario representation with their respective semantic segments as defined by OpenScenario's specification. 

This study introduces a priority-based assembling architecture (\textbf{Algorithm \ref{algo1}}) tailored for static element matching and dynamic element concatenation between scenario representations and DSL corpus fragments. Commencing with the initialization of a DSL material library -- the product of meticulous manual curation -- this repository serves as a resource pool for populating target DSL-based files. It comprises essential event fragments such as acceleration, deceleration, lane-changing, and synchronization maneuvers amongst others \cite{asam_openscenario_2021}. The scenario representation, standardized into a dictionary-format result (\textit{e.g.}, \textit{.json}), is derived from the LLM-powered parser as expounded upon in Section \ref{subsec3.3}. An action chain is an isolated sequence of maneuvers performed by traffic entities. The objective is to delineate each standard action originating from a rudimentary scenario representation of background traffic, thereby crafting an interpretable, finely-tuned DSL control sequence. Take, for instance, the lateral movement component within a traffic participant's scenario representation labeled as "overtaking". Such a label, while conceptually concise, encompasses a suite of discrete standard DSL actions -- essentially, "change lanes - accelerate - change lanes - continue at speed" -- which collectively constitute the complex maneuver traditionally recognized as "overtaking".

\begin{algorithm}
\caption{ Priority Ranking-based DSL Padding}\label{algo1}
\begin{algorithmic}[1]
\Require DSL material library $\mathcal{L}$, target DSL-based file $\mathcal{T}$, LLM scenario representation $\mathcal{R}$, action chain of traffic participants $\mathcal{A}$

\Comment{\textcolor{blue}{DSL Material Library Filling}}
\State $\mathcal{L}\leftarrow\textbf{Simulator Element PriorKnowledge}$
\State $\mathcal{L}\leftarrow\textbf{Route Random Search}$

\Comment{\textcolor{blue}{Static Element Matching}}
\State$\mathcal{T}.\textbf{Climate}^{\prime}\xleftarrow{\mathcal{R}.Climate}\mathcal{L}$
\State$\mathcal{T}.\textbf{Topology}\xleftarrow{\mathcal{R}.Topology}\mathcal{L}$
\State$\mathcal{T}.\textbf{TransportationFacilities}\xleftarrow{\mathcal{R}.Transportation Facilities}\mathcal{L}\sim\mathcal{T}.\textbf{Topology}$
\State$\mathcal{T}.\textbf{TemporaryChanges}\xleftarrow{\mathcal{R}.Temporary Changes}\mathcal{L}\sim\mathcal{T}.\textbf{Topology}$
\State$\mathcal{T}.\textbf{EgoVehicle}\xleftarrow{\mathcal{R}.Ego Vehicle}\mathcal{L}\sim\mathcal{T}.\textbf{Topology}~\textbf{\&}~\mathcal{T}.\textbf{TemporaryChanges}$
\State$\mathcal{T}.\textbf{TrafficParticipants}.(\textbf{type}\&\textbf{position\_} \textbf{relation})\xleftarrow{\mathcal{R}.TrafficParticipants}\mathcal{L}\sim\mathcal{T}.\textbf{Topology}~\textbf{\&}~\mathcal{T}.\textbf{EgoVehicle}$

\Comment{\textcolor{blue}{Dynamic Element Concatenating}}
\State$\mathbb{A}\leftarrow\text{LLM powered action-chain decomposition}$
\For{$a$ in $\mathbb{A}$}
    \State$\mathcal{A}\xleftarrow{\mathcal{R}.TrafficParticipants.oracle}a$
\EndFor
\State$\mathcal{A}\xleftarrow{\mathcal{R}.TrafficParticipants.global~behavior}\mathcal{A}.append(\mathcal{L}.autopilot())$
\State $\mathcal{T}\leftarrow\mathcal{T}+\mathcal{L}(\mathcal{A})$

\State \textbf{return} $\mathcal{T}$ 
\end{algorithmic}
\end{algorithm}

\begin{figure}[t!]
\centering
\includegraphics[width=13cm]{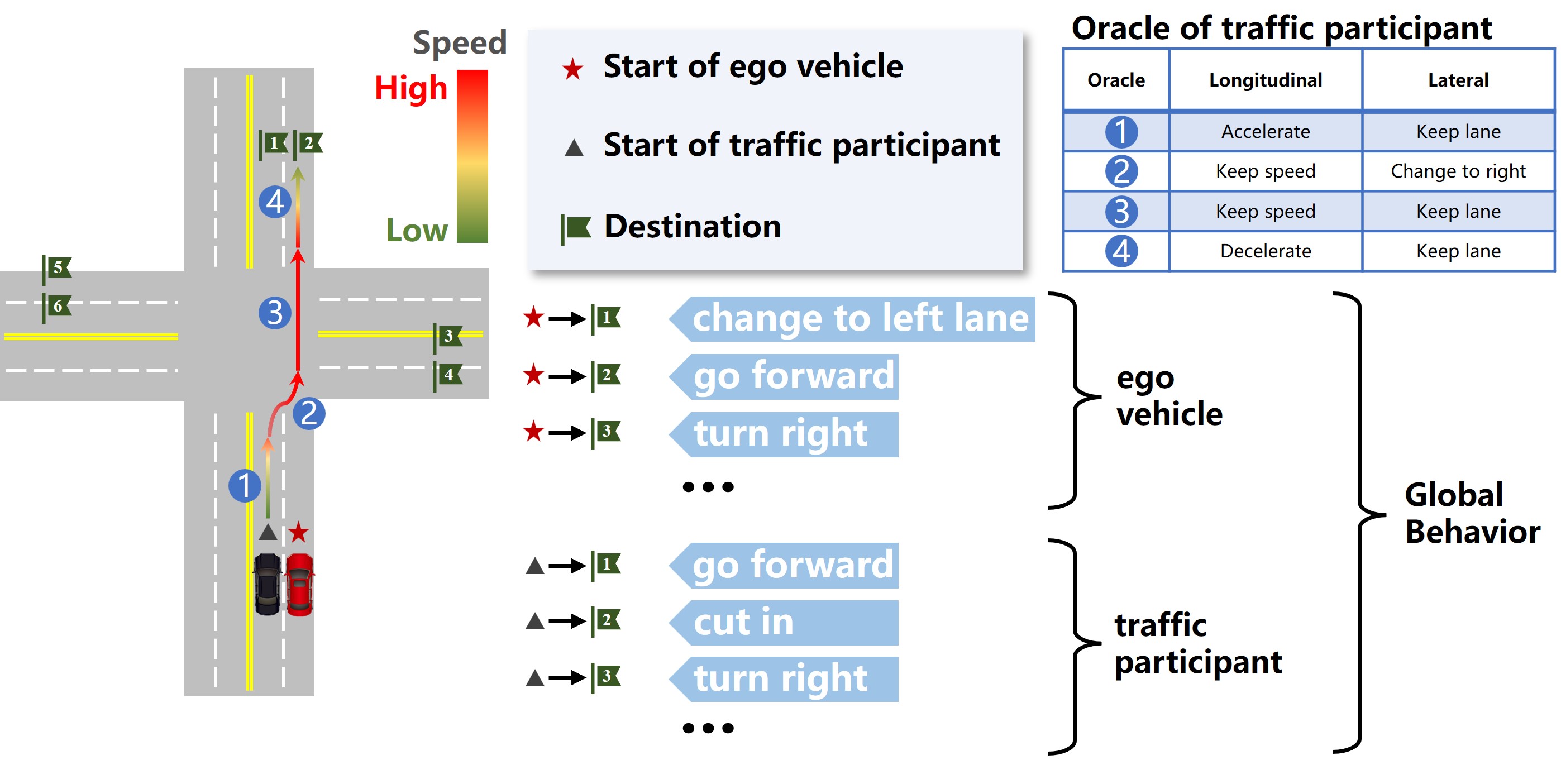}
\caption{Schematic diagram and labels for driving route search at intersection. The \textcolor{dateplum}{$\bigstar$} and \textcolor{deepgreen}{$\blacktriangle$} symbolize the initial positions of the ego vehicle and traffic participant, respectively, while the green flag marks potential endpoints. Task labels vary from the starting point to different endpoints, facilitating scenario search matching. For instance, a traffic participant reaching endpoint 2 can be categorized into four oracle stages (indicated by the circle's number) for DSL, both longitudinally and laterally.}
\label{fig_route_search.jpg}
\vspace{-0cm}
\end{figure}

Subsequent to data acquisition from the scenario simulator, the repository necessitates enrichment (refer to lines 1-2). This involves the assimilation of prior knowledge concerning simulator components, for instance, APIs for weather configuration and the selection of scenery maps. The objective is to supply route-based candidates for evaluation involving both ego vehicles and background traffic. This is realized by stochastically generating vehicular pathways -- ranging from start to target points -- within the set of feasible waypoints provided by the simulator, and annotating these points with precision. As illustrated in Fig.\ref{fig_route_search.jpg}, consider a scenario situated within a two-lane junction. It is necessary to assign the randomly generated origin and destination coordinates with relevant labels indicating the intended interaction patterns with the ego vehicle. Each pair of route endpoints, along with their associative labels, is to be integrated into the repository, paving the way for further synergistic correlation with the key-value pairs of the scenario representation.

Subsequently, the DSL-based file designated for scenography was methodically populated with corresponding semantic units to articulate the scenario (refer to lines 3-14). The process commenced by sorting the static elements based on priorities within the hierarchical scenario repository, thereby ensuring congruity in element matching and averting redundant designations (as outlined in lines 3-8). For example, when "intersection" is delineated within $\langle\mathrm{road~topology}\rangle $, the $\langle\mathrm{road~marker}\rangle$ ought not to be signified as "broken lane," given that vehicular lane transitions at intersections are typically regulated to "solid lane" in compliance with traffic laws. Any contradictions arising in the scenario paperwork would suggest either a faltering in LLM's parsing or an erratum within the scenario's linguistic text, meriting manual scrutiny for clarification. Proceeding from lines 3 to 8, the protocol harnesses the precedence of the static hierarchy; materials are selectively retrieved from the library and embedded into the DSL-based file as delineated by the scenario narrative. The higher the hierarchical standing, the earlier its match is sought. Climate and Topology are bestowed the pinnacle of precedence. Illustrated by the expression $\mathcal{T}.\textbf{TransportationFacilities}\xleftarrow{\mathcal{R}.TransportationFacilities}\mathcal{L}\sim \mathcal{T}.\textbf{Topology}$, it is interpreted as deriving the pertinent fragment from $\mathcal{L}$ to the key within $\langle\mathrm{road~topology}\rangle $ and $\langle\mathrm{traffic~sign}\rangle $ located in $\mathcal{R}$. The parameters of these segments are then calibrated, or a match is sought with a corresponding fragment, predicated on the value associated with the key. And the fragment elected based on $\mathcal{L}$ is subsequently transcribed to the correlating site within $\mathcal{T}$. The symbol "$\sim$" signifies adherence to a higher hierarchical parameter within the $\mathcal{T}$ 's Topology. 

For the residuum encompassing dynamic elements such as $\textbf{TrafficParticipants.(oracle \& global behavior)}$, an approach deploying action dissection and concatenation is utilized, as detailed from line 9 to 14. Fig.\ref{fig_route_search.jpg} delineates an instance where the "cut in" maneuver executed by black traffic vehicles in relation to the ego vehicle can be resolved into a sequence of interconnected actions, typified by "acceleration, right-lane change, cruise, deceleration". This sequenced behavior is further deconstructed into a continuum of action chains through the use of LLM, aligning each with their respective, intricate event fragments residing in the library. Post-completion of the action chain, in scenarios where the vehicle is yet to arrive at its target terminus, the automated navigation algorithm embedded within the action chain -- as described in line 13 -- dictates an appendage from $\mathcal{L}$, propelling the vehicle to fulfill the comprehensive task. To encapsulate the process, the specific $\mathcal{L}(\mathcal{A})$ fragment, representing the action chain, is integrated into the designated locus within $\mathcal{T}$.

\subsection{Evaluation Metrics for System under Test}\label{subsec3.5}

Appropriate evaluation metrics can be incorporated into the target DSL-based file to activate the simulator's monitoring system, which persistently scrutinizes the SUT behavior on a frame-by-frame basis throughout the entire execution of the scenario. These evaluative indicators encompass:

\begin{itemize}
\item { {\textit{Rule violation.}} This metric ascertains if the SUT contravenes traffic regulations, which entails assessments such as running stop test, running red light test, wrong lane test, and lane-keeping test. For instance, a lane invasion sensor could be configured within the Carla simulation for vigilant surveillance.}

\item {{\textit{Collision.}} This metric determines if the SUT has experienced any collisions with other vehicles or pedestrian objects in its vicinity. Implementing a collision detection sensor within Carla would facilitate perception and aid in recording such events. }

\item {{\textit{Time out.}} This metric gauges the SUT's ability to navigate to the proximity of the designated endpoint within the stipulated maximal time limit. This timeframe is adaptable and is predicated upon the length of the ADS route as well as the velocity constraint factors inherent within the scenario.}

\end{itemize}

\section{Simulation Experiment}\label{sec4}

In this section, we empirically validate the advantages of the T2S framework by leveraging extensive textual inputs and conducting well-structured simulation experiments. The implementation utilizes the ASAM OpenScenario standard format \cite{asam_openxontology_2021}. The simulations are conducted using the Carla \cite{dosovitskiy2017carla} simulator, and a customized DSL file parser is integrated into our optimized Carla scenario execution environment.

\subsection{Research Questions}\label{subsec4.1}

To ascertain the efficacy of the T2S framework, we designed three research questions (RQs) targeting key aspects of scenario generation performance. RQ1 examines the extent to which the LLM-powered testing text parsers align with the granularity of different scenario groundtruth. RQ2 investigates the fidelity with which T2S reproduces scenarios from testing description texts. RQ3 explores the effectiveness of the T2S evaluation mechanism in assessing the driving conduct of diverse SUTs. 

\begin{itemize}
\item { \textbf{RQ1: How effectively does T2S's testing text parser perform at capturing the nuances across various scenario complexity levels?} }

\item { \textbf{RQ2: Is T2S capable of accurately translating comprehensive testing description texts into executable scenarios?} }

\item { \textbf{RQ3: Does T2S facilitate precise evaluation of the driving behaviors for various of SUTs?} }

\end{itemize}

\subsection{Benchmark}\label{subsec4.2}

\subsubsection{Testing Text Benchmark}\label{subsec4.2.1}

To ensure the scientific rigor of the testing text inputs, we have selected three exemplary scenario description languages that are illustrative of the industry standards: (1) The pre-collision scenario descriptions outlined by the National Highway Traffic Safety Administration (NHTSA) \cite{najm_pre-crash_2007}; (2) The management regulations released in the Active Safety Advanced Driver-Assistance Systems (ADAS) Test Protocol, 2024 Edition, by the China New Car Assessment Program (C-NCAP) \cite{cncap_adastestprotocol_2024} and (3) customized natural language texts for testing scenarios, meticulously crafted by testing experts in the field.

It is important to emphasize that not every test scenario provided by the NHTSA and C-NCAP is applicable to our research. Scenarios that did not pertain to the evaluation of ADS performance or that were incompatible with simulation capabilities were omitted from the study. After a careful selection process, a total of 92 test texts were retained for our analysis (NHTSA: 23; C-NCAP: 17; Customized: 52).

\subsubsection{Test Scenario Benchmark}\label{subsec4.2.2}

Owing to the versatile mapping of a single testing text to various locales within the diverse map topographies in Carla, a solitary test narrative is capable of spawning multiple analogous scenarios. Utilizing the 92 curated texts, we actualized the generation of 368 test scenarios, encompassing all six intrinsic road environments within Carla, which span urban, suburban, and rural settings.

\subsubsection{Systems under Test}\label{subsec4.2.3}

We selected four distinct types of AD stacks as SUTs, that is, Autoware \cite{kato2018autoware}, Apollo \cite{apolloauto2021}, Interfuser \cite{shao2023safety}, Dora-RS\footnote{\href{https://github.com/dora-rs/dora}{https://github.com/dora-rs/dora.}}. Autoware and Apollo are recognized as the most sophisticated and widely-adopted application-level AD systems within the industry. Interfuser, developed by OpenDILab, is an advanced end-to-end ADS based on sensor fusion\footnote{\href{https://leaderboard.Carla.org/leaderboard/}{https://leaderboard.Carla.org/leaderboard/.} Interfuser ranks first in the Carla Leaderboard.}. Doara-RS presents a revolutionary robotic framework, infusing modernity into robotic applications; it is a prototype that hinges on a purely visual modality for its AD capability and boasts a more flexible communication framework than ROS2\footnote{\href{https://github.com/ros2/ros2}{https://github.com/ros2/ros2.}}. \textcolor{black}{Note that all SUT entities are jointly simulated via the official or third-party Carla bridge. The SUTs manage the ego vehicle through various interfaces, while test scenarios are governed by decoding a DSL script.}

\subsubsection{LLMs used in Text Parse}\label{subsec4.2.4}
Heterogeneous LLMs: For the task of test text parsing, we employed three principal LLMs, including Yi-34B-Chat \cite{young2024yi}, OpenAI GPT-3.5 \cite{openai_chatgpt_2022}, GPT-4 \cite{openai_chatgpt_2022}. In alignment with the protocol delineated in Section \ref{subsec3.3} (refer to Fig.\ref{fig_engineering}), the deployment of these LLMs was facilitated through their respective official APIs, ensuring systematic invocation of the LLM agents for parsing responses. 

Ablation study: To elucidate the significance of each module in the parser, ablation studies were conducted, systematically omitting individual modules to assess their impact. The complete sequence of operations for the parser: Basic Prompt (BP) → Few-Shot (FS) → Chain-of-Thought (CoT) → Syntax Alignment Checking (SAC) → Self-Consistency (SC), enabled the creation of five distinct parser configurations for the purpose of ablation studies. These configurations are denoted as LLM-BP, LLM-BP-FS, LLM-BP-FS-CoT, LLM-BP-FS-CoT-SAC, and LLM-BP-FS-CoT-SAC-SC respectively. 

\subsection{Experiment Settings}\label{sec4.3}
\subsubsection{RQ1.The matching degree across the element-level of detail with the LLM-powered parser}\label{sec4.3.1}

To appraise the hierarchical parsing capabilities of our testing scenario parser -- specifically the degree of concordance between scenario representations and the decomposed elements of the testing text -- a comparative analysis between the generated representations and the established groundtruth is imperative. In exploring the generation of virtual testing scenarios, we have examined cutting-edge methodologies recently introduced, such as RMT \cite{deng2021rmt} and TARGET \cite{deng2023target}, for text-based scenario generation. RMT was deemed incongruent as a baseline methodology since it is limited to the generation of video-based open-loop test scenarios, without the provision for a closed-loop feedback model essential for virtual simulation testing. Conversely, TARGET, with its focus solely on parsing traffic rule inputs, fell short of the requisites for crafting the comprehensive virtual scenario testing scripts central to our evaluation framework. 

\subsubsection{RQ2. The effectiveness of T2S in generating scenarios -- feasibility, accuracy, and efficiency}\label{sec4.3.2}

To gauge T2S's impact on mitigating the manual labor involved in constructing testing scenarios, we enlisted four ADS testing experts to juxtapose human efforts with different LLM-based implementations of T2S, thereby providing a comprehensive assessment of their feasibility, accuracy, and efficiency. 

For the evaluation of the feasibility of DSL files, we implemented a monitor to track and log any read or run-time errors that may occur in the generated OpenScenario DSL files. 

To appraise the accuracy of the scenarios generated by T2S, we recruited 15 automotive and transportation students to manually assess the congruity between the recorded output videos of the scenarios and their corresponding textual descriptions. They quantified the accuracy of T2S's scenario generation by attributing a match score ranging from 0 to 1 -- using a 0.1 discrete interval scale, with higher scores indicating a greater match. 

Concurrently, the efficiency of T2S’s rapid and precise automated testing scenario generation was evaluated by measuring the time taken for scenario construction. 

\subsubsection{RQ3. Compatibility of T2S evaluator with heterogeneous SUTs}\label{sec4.3.3}

All 4 varieties of SUTs have been seamlessly integrated to ensure compatibility with the DSL testing files. The configuration for these SUTs is restricted to designating starting and ending points, deliberately precluding human intervention in modulating any intermediate behaviors. Throughout the execution of each scenario, the performance of the SUT is attentively observed, with the objective of recording its evaluation metrics. The data accrued from these observations is then utilized to synthesize comprehensive test reports.

\subsection{Evaluation Metrics}\label{sec4.4}
\subsubsection{RQ1. The matching degree across the element-level of detail with the LLM-powered parser}\label{sec4.3.1}

To quantify the LLM-powered parser's proficiency in element-level analysis, the correctness of parsing for each constituent sub-element within the scenarios was methodically evaluated. The parser's proficiency at managing and interpreting intricate textual information is reflected in the ratio of accurately parsed elements across the entirety of the test cases. This metric serves as an indicator of the parser's overall analytical capability.

\subsubsection{RQ2. The effectiveness of T2S in generating scenarios - feasibility, accuracy, and efficiency}\label{sec4.3.2}

We quantify the capabilities of T2S and human testing experts across three primary dimensions: feasibility, accuracy, and efficiency. Concerning \textbf{feasibility}, the incidence of read errors and run-time errors throughout the scenarios serves as a crucial benchmark for ascertaining the executability level of T2S. The lower, the better.
With respect to \textbf{accuracy}, we tasked judges with appraising two critical aspects: 1) Semantic fidelity (0-1) -- this evaluation gauges the semantic coherence between the input description and the generated scenario; and 2) Driving rationality (0-1) -- this evaluation determines the extent to which the traffic participants' behavior in the generated scenario aligns with normal traffic patterns.
Regarding \textbf{efficiency}, we elected to utilize the metric of scenario construction time as a basis for comparison with the proficiency and pace of professional human testing experts in scenario creation. 

\subsubsection{RQ3. Compatibility of T2S evaluator with heterogeneous SUTs}\label{sec4.3.3}

Throughout the execution phase of the scenario, three critical indicators are selected to assess the competency of the AD stack: rule violation, collision, and timeout. For the timeout criterion, a threshold is established by dividing the total distance of the test route by 10\% of the speed limit, which serves as the maximum allowable time for the AD stack to complete the test route.

\section{Results}\label{sec5}
\subsection{RQ1. The matching degree across the element-level of detail with the LLM-powered parser}\label{sec5.1}

Tab.\ref{tab:accuracy_element} delineates the parsing precision of three different LLMs for each element when a comprehensive prompt pipeline is employed, encompassing a total of 92 language descriptions within this study. The accuracy of the scenario representations produced by the LLM-based testing text parser is contrasted with the groundtruth labeled by human experts within the original scenario descriptions. This comparison evaluates their proficiency in comprehending and extracting knowledge from textual language. ChatGPT-4.0 exhibits superior language comprehension abilities relative to Yi-34B-Chat and ChatGPT-3.5, as evidenced by its closer alignment with human-level accuracy in interpreting most scenario elements, attributable to its more advanced pre-trained model. While Yi-34B-Chat outperforms ChatGPT-4.0 in parsing certain elements slightly, it maintains a generally stronger performance than ChatGPT-3.5.

\begin{table}[!t]
\centering
\caption{The accuracy of hierarchical scenario element parsing for different LLMs based parsers (368 scenarios). RT-Road Topology, TF-Transportation Facilities, TC-Temporary Changes, TP-Traffic Participants, C-Climate, EV-Ego Vehicle.}
\label{tab:accuracy_element}
\begin{tabular}{>{\raggedright}p{0.25\linewidth}>{\raggedright\arraybackslash}p{0.18\linewidth}>{\raggedright\arraybackslash}p{0.18\linewidth}>{\raggedright\arraybackslash}p{0.18\linewidth}}
\toprule
\multicolumn{1}{c}{\textbf{Element}} & \textbf{Yi-34B-Chat} & \textbf{ChatGPT-3.5} & \textbf{ChatGPT-4.0} \\
\midrule
RT.$\langle\mathrm{topology}\rangle$ &\textbf{0.96} & 0.89 & 0.95 \\
RT.$\langle\mathrm{lanes}\rangle$ & 0.96 & 0.85 & \textbf{0.98} \\
TF.$\langle\mathrm{road marker}\rangle$ & 0.90 & 0.90 & \textbf{0.97} \\
TF.$\langle\mathrm{traffic sign}\rangle$ & 0.99 & 0.97 & \textbf{1.00} \\
TC.$\langle\mathrm{type}\rangle$ & 0.83 & 0.80 & \textbf{0.95} \\
TC.$\langle\mathrm{position relation}\rangle$ & 0.81 & 0.76 & \textbf{0.92} \\
TP.$\langle\mathrm{type}\rangle$ & 0.95 & 0.89 & \textbf{0.98} \\
TP.$\langle\mathrm{position relation}\rangle$ & 0.83 & 0.62 & \textbf{0.83} \\
TP.$\langle\mathrm{longitudinal oracle}\rangle$ & 0.71 & 0.59 & \textbf{0.74} \\
TP.$\langle\mathrm{lateral oracle}\rangle$ & 0.72 & 0.53 & \textbf{0.77} \\
TP.$\langle\mathrm{global behavior}\rangle$ & \textbf{0.86} & 0.70 & 0.83 \\
C.$\langle\mathrm{type}\rangle$ & 0.98 & 0.91 & \textbf{0.99} \\
C.$\langle\mathrm{density}\rangle$ & 0.87 & 0.68 & \textbf{0.97} \\
C.$\langle\mathrm{time}\rangle$ & 0.90 & 0.86 & \textbf{0.95} \\
EV.$\langle\mathrm{type}\rangle$ & 0.99 & 0.98 & \textbf{0.99} \\
EV.$\langle\mathrm{position}\rangle$ & \textbf{0.95} & 0.90 & 0.95 \\
EV.$\langle\mathrm{global behavior}\rangle$ & 0.82 & 0.76 & \textbf{0.86} \\
\rowcolor{gray!25} 
Average & 0.88 & 0.80 & \textbf{0.92} \\
\bottomrule
\end{tabular}
\end{table}

Within the scope of LLM analysis, explicit static expressions such as road topology, climate, and vehicle types are typically comprehended with relative ease. However, LLMs often struggle with the interpretation of complex vehicular behaviors and the nuances of multiple traffic participants' interactions, particularly when dealing with implicit expressions. For instance, the statement "\textit{The ego vehicle encounters a vehicle cutting into its lane from a lane of static traffic}" can be misinterpreted by an LLM, which might fail to infer the implication of numerous stationary vehicles present in that lane, sometimes only generating a single vehicle in response. This issue is frequently linked to the error propagation inherent in the input-output examples used during the few-shot, as we have identified that LLMs tend to yield correct results when provided with similar instances. Moving forward, our research will delve into refining prompt engineering techniques to foster a more encompassing comprehension of traffic components by LLMs. 

In a horizontal comparative analysis among the three LLMs, ChatGPT-4.0 holds an average parsing accuracy of 92.3\%, which surpasses Yi-34B-Chat and ChatGPT-3.5 by 4.2\% and 11.9\%, respectively, standing at 88.1\% and 80.4\%. This advantage is particularly noticeable in understanding dynamic behaviors within oracles, preventing significant performance degradation. Looking ahead, enhancing LLM parsing capabilities for certain concealed elements could benefit from a multimodal data analysis approach that integrates text, imagery, and video elements.

\subsection{RQ2. The effectiveness of T2S in generating scenarios -- feasibility, accuracy, and efficiency}\label{sec5.2}
\subsubsection{Feasibility}\label{sec5.2.1}

\begin{figure}[t!]
\centering
\includegraphics[width=11cm]{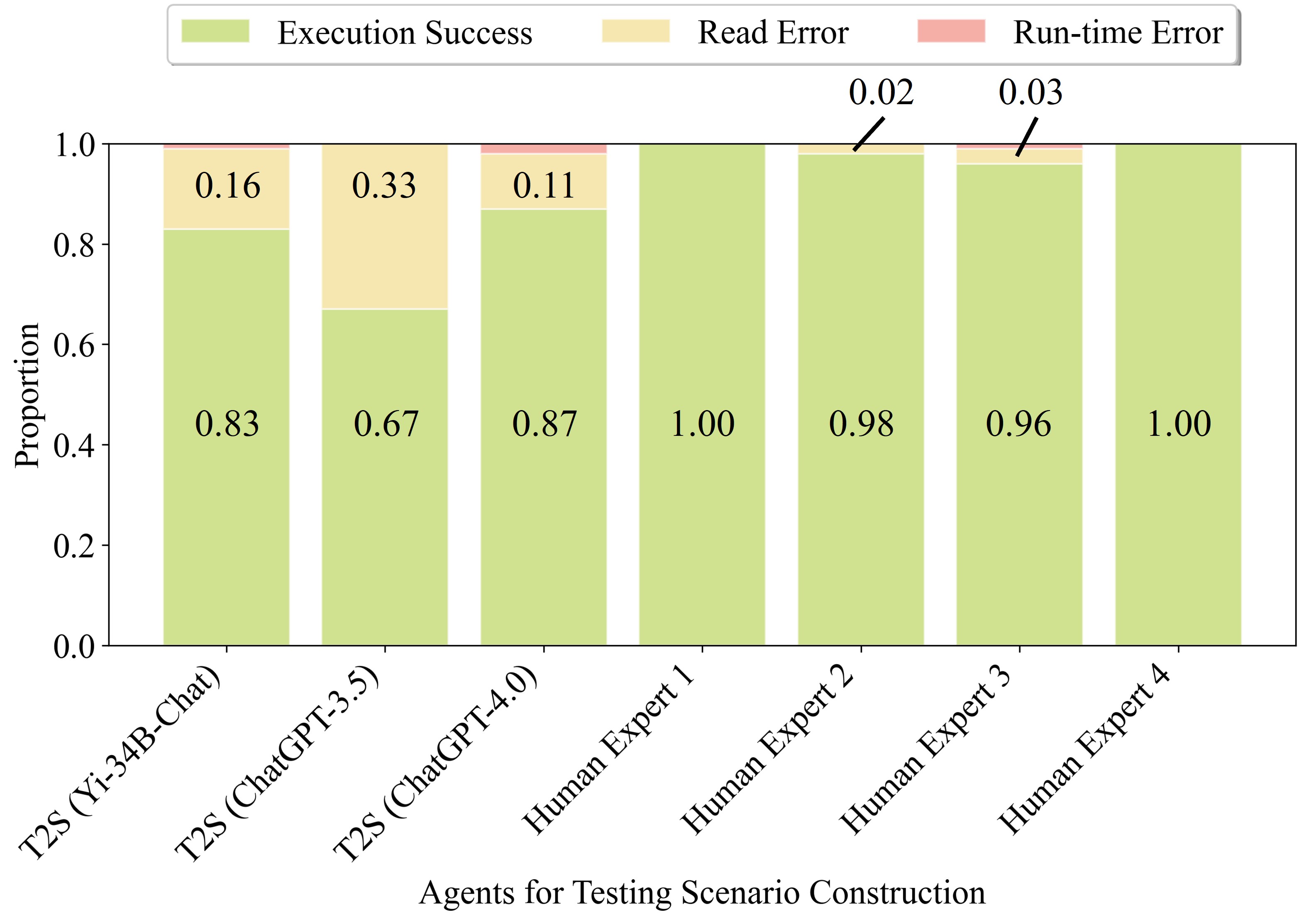}
\caption{The statistic proportion of DSL-file made by different LLMs-based T2S and human expert agents.}
\label{fig_proportion}
\vspace{-0cm}
\end{figure}

In our assessment, T2S-generated scenarios based on different LLMs have a markedly lower success rate when compared to human experts, as elucidated in Fig.\ref{fig_proportion}. Human experts achieved near-perfect rates of successful script formation. This is because there is no temporal restriction on scenario construction -- experts can repeatedly revise the scripts through an iterative process of reading, comprehending, authoring, debugging, testing, and modifying until they achieve success. In contrast, LLM-based T2S lacks the capability to verify the feasibility of their generated scenario representations during the construction process. We have, however, preserved the adaptable operation space of
LLM-based systems, aspiring to the principle of meticulous understanding over a rigid, hard-coded approach. Drawing insights from the human scenario construction methodology, our future research aims to improve the generation of T2S through a closed-loop feedback mechanism.

ChatGPT-4.0 maintains a significant advantage over its counterparts regarding the creation of executable scenarios, securing an 87.31\% success rate, which stands considerably higher than the 5.47\% and 20.21\% achieved by Yi-34B-Chat and ChatGPT-3.5, respectively. Our analysis uncovered that read errors typically arise when LLMs select components that fall outside the established dataset, lacking corresponding DSL fragments, which may hinder the generation of accurate DSL files. Runtime errors commonly occur due to abrupt terminations in the runtime, associated with legitimate but imperceptibly contradictory parameter configurations within the DSL text of the test scenario. Human experts, with their extensive experience in fine-tuning scenario parameters, are adept at averting such issues.

\subsubsection{Accuracy}\label{sec5.2.2}

For the purpose of evaluating the accuracy of the scenario parser, we utilized the top-performing ChatGPT-4.0. A sampling of 23 scenarios out of the 368 was randomly selected, and human evaluators were invited to rate the scenarios for semantic fidelity and driving rationality. The distribution of these scores is depicted in Fig.\ref{fig_score_distribution}, with further details provided in the supplementary materials\footnote{\href{https://docs.google.com/forms/d/e/1FAIpQLSdUoDLjNLBwgTNYyCRd9KYRjg4RML9LKa-GXPeeKcCOjiSk5g/viewform?usp=sf_link}{https://docs.google.com/forms/d/e/1FAIpQLSdUoDLjNLBwgTNYyCRd9KYRjg4RML9LKa-GXPeeKcCOjiSk5g/viewform?usp=sf\_link.}}. Acknowledging the inherent subjectivity and variability among different human judges -- some being more inclined to award higher scores while others favor lower scores for the same item -- we employed the Intraclass Correlation Coefficient (ICC) bidirectional random effects model \cite{shrout1979intraclass}. This model serves to assess the consistency or reliability of quantitative data acquired through repeated measures, including multiple evaluators scoring the same variables. The statistical analysis revealed that the ICC value for semantic fidelity stands at 0.92, denoting superb consistency. Meanwhile, the ICC for driving rationality is 0.76, indicative of good consistency. The comparatively lower value for the latter can be ascribed to the diverse driving experiences of the evaluators, which in turn influence their perceptions and interpretations of traffic behavior. For instance, seasoned drivers might anticipate that the conduct of road users should account for interaction with other participants -- a factor not encapsulated in the original DSL scenario execution file. Consequently, in certain scenarios, some traffic participants might display driving behaviors that are deemed illogical by these evaluators.

\begin{figure}[t!]
\centering
\includegraphics[width=13cm]{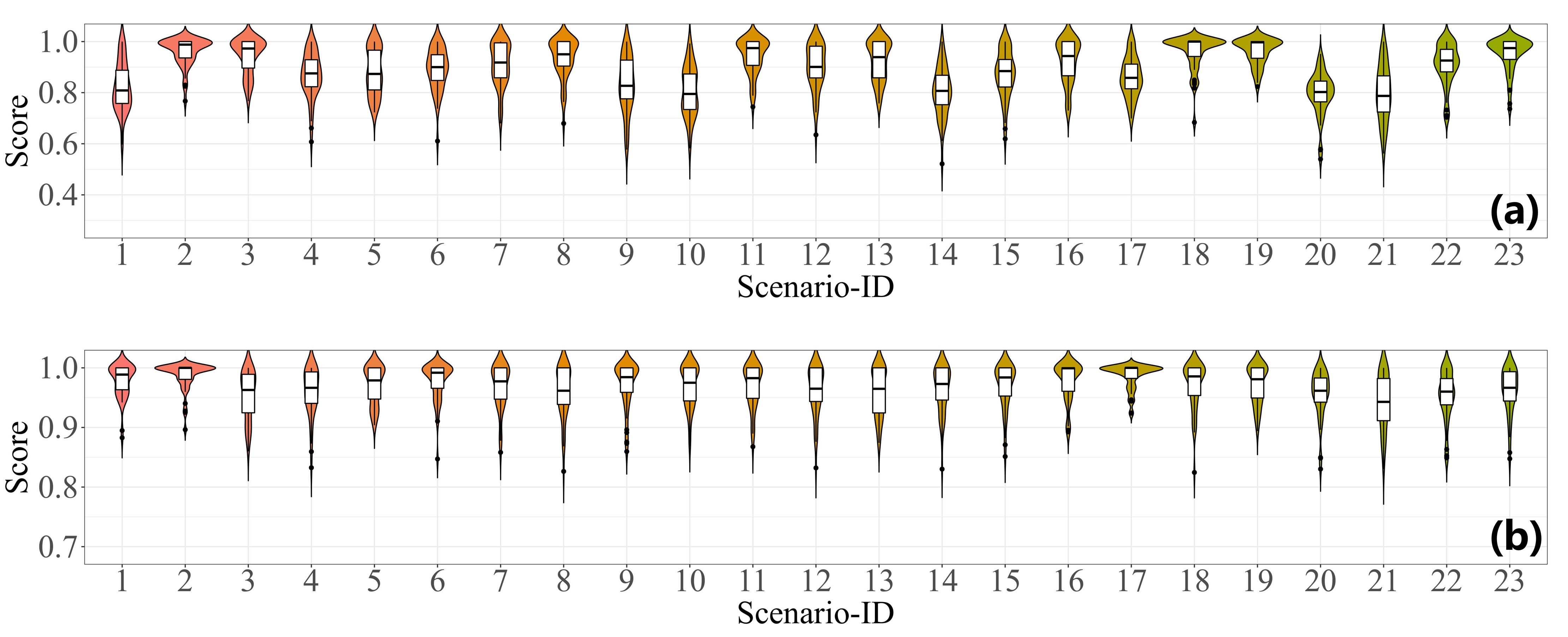}
\caption{The scoring distribution of (a) semantic fidelity and (b) driving rationality of 23 generated scenarios from T2S by human judges.}
\label{fig_score_distribution}
\vspace{-0cm}
\end{figure}

Our analysis extended into the exploration of factors contributing to the lower scores assigned to certain scenarios. It appears that the deficiency in the scenario's graphical representation, particularly the inadequate depiction of water stains on the roadway (as with scenario ID 1 in subset (a)), coupled with the omission of road friction coefficients, resulted in an unrealistic portrayal of the expected loss of control. Similarly, in scenario ID 23 in subset (b), evaluators frequently noted a discrepancy between the exhibited traffic flow behaviors and standardized norms. Notwithstanding these instances, the average scores dispensed by most evaluators for both semantic fidelity and driving rationality were predominantly high, averaging 0.97 and 0.95, respectively. This affirmatively validates the capability of T2S in generating scenarios that are both semantically accurate and adherent to rational driving expectations, underscoring its utility in the intended application domain.

\subsubsection{Efficiency}\label{sec5.2.3}

\begin{figure}[t!]
\centering
\includegraphics[width=8cm]{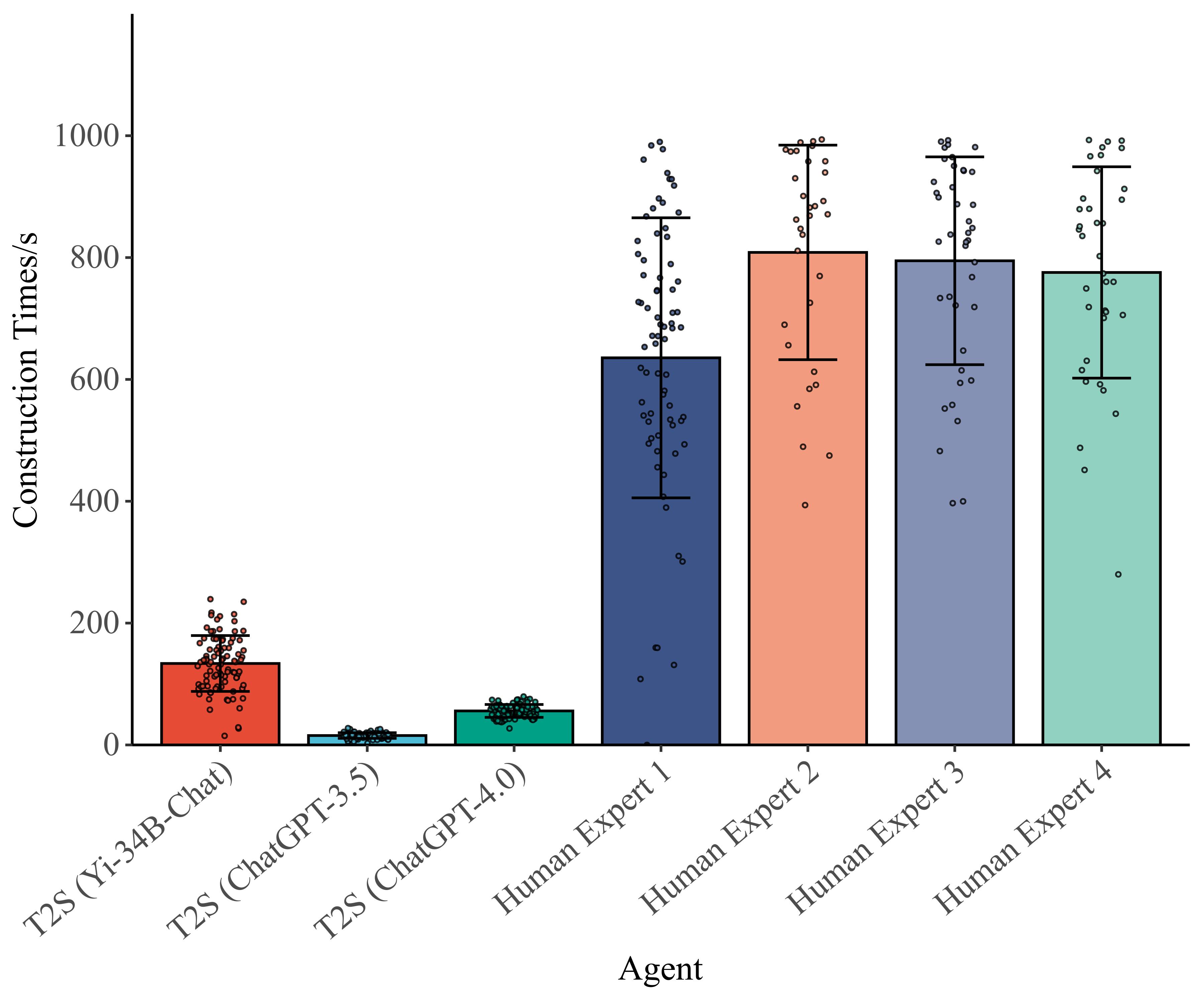}
\caption{Bar graph with error bars of construction time of T2S and human expert agents.}
\label{fig_bar}
\vspace{-0cm}
\end{figure}

To assess the impact of the T2S framework on lessening the intensity of human labor, we quantified the scenario construction duration for various LLM-based T2S agents and a cohort of four human experts (4 AD testing engineers, of which three have 2 years of work experience and one has more than 4 years). Demonstrated in Fig.\ref{fig_bar} as a bar graph with error bars, the results indicate that T2S delivers a marked advantage in terms of efficiency, with construction times consistently falling below 200 seconds. Predominantly, time expenditure is attributed to LLM's cloud-based inference, where the ChatGPT-X series excels in response agility. Comparatively, the mean construction times for ChatGPT-3.5 and ChatGPT-4.0 stand at approximately 15 seconds and 55 seconds, respectively. The Yi-34B-Chat exhibits a longer average construction time, which is around 130 seconds. In stark contrast, the construction time for human experts notably exceeds that of T2S agents, with an average duration surpassing 600 seconds, compounding greater variability in scenario construction times. Although the efficiency of human professionals tends to improve over multiple attempts, achieving the stable processing time exhibited by T2S agents is challenging. Human participants are susceptible to grammatical and detailed errors, necessitating frequent iterative cycles of verification and debugging. In summary, the T2S framework greatly mitigates the time invested in constructing virtual scenarios, thereby diminishing the extent of human labor required.

\subsection{RQ3. Compatibility of T2S evaluator with heterogeneous SUTs}\label{sec5.3}

\begin{figure}[t!]
\centering
\includegraphics[width=13cm]{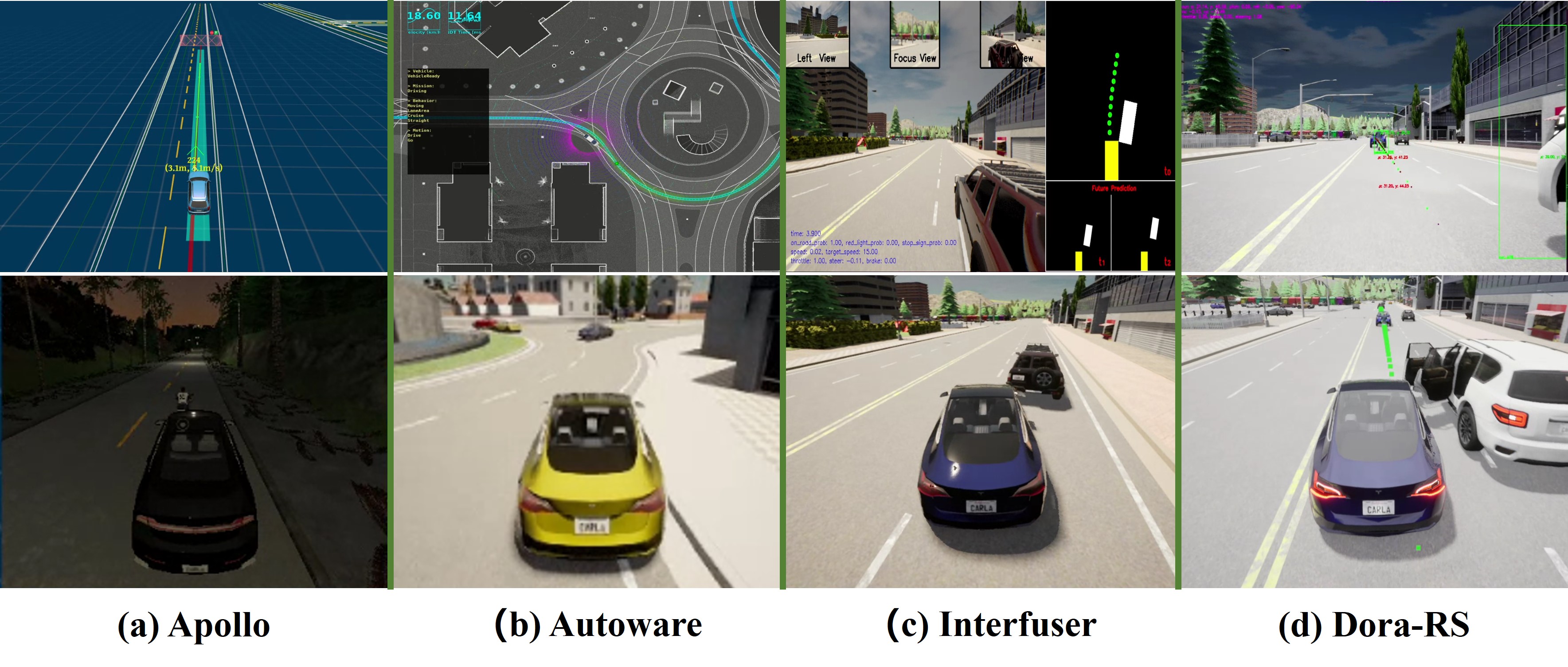}
\caption{Snapshots of violation scenarios for four different SUTs. (a) Apollo. Chasing a two wheeled motorcycle with no lights on in the dim dusk; (b) Autoware. Violating the right solid line when entering the roundabout; (c) Interfuser. When changing lanes and overtaking, it collides with the rear end of the same lane; (d) Dora RS. The sudden opening of the door by a parked vehicle on the roadside caused a collision.}
\label{fig_snapshot}
\vspace{-0cm}
\end{figure}

Fig.\ref{fig_snapshot} delineates the snapshots of testing scenarios for four divergent SUTs -- Apollo, Autoware, Interfuser, and Dora RS. The upper portion of the figure emanates from the SUT systems' developer-provided visual interfaces, while the lower segment illustrates the controlled ego vehicle within the Carla. The OpenScenario execution files, paramount for these tests, have been meticulously refined to ensure cross-compatibility with the aforementioned SUT variants. Detailing the SUTs' performance, Apollo experienced a collision, attributable to its deficient recognition capabilities in detecting two-wheeled motorcycle. Autoware, faced with an abruptly reducing target point in Carla's roundabout representation, deviated onto a high curvature trajectory, ultimately transgressing the lane boundary. Interfuser's inadequacy surfaced during a lane-change maneuver involving close-quarters interactions, where it failed to maintain a precise account of its dimensions, leading to minor abrasions. Dora-RS's perceptual mechanisms faltered, lacking the requisite training to detect out-of-distribution, such as road-side vehicles opening their doors, culminating in a failure to recognize the hazard and consequently a severe collision.

Tab.\ref{tab:accuracy_element} meticulously documents the number of safety violation occurrences for each SUT within the extensive ensemble of 368 test scenarios. Apollo and Autoware have displayed commendable rule adherence and reduced collision instances, a testament to their elaborate compliance state machines and robust security frameworks that command their behavior. In stark contrast, Interfuser has been identified as the preeminent transgressor with a total of 76 rule violations, leading the cohort in this regard. Concurrently, Dora-RS held the undesirable distinction of being the most collision-prone SUT, recording 67 incidents. Analysis attributes this vulnerability to the frequent lapses in its vision-only detection system, which transmits flawed perceptual data, predisposing the system to engage in rear-end collisions or margin-centric incidents evocative of the scenarios delineated in Figure 14 (d). A peculiarity was noted in the comparison between Apollo and Autoware's timeout occurrences: Apollo presented with a surprisingly elevated rate, disparate from its projected operational capabilities. Investigations attributed this divergence to potential disparities within Guardstrikelab's Apollo-Carla-Bridge software parameters, which may sporadically lose connection, thereby precipitating signal interruption, collisions, or even inducing system inoperability. These findings have been corroborated through discussions within the open-source community forums\footnote{\href{https://github.com/guardstrikelab/carla_apollo_bridge/issues/159}{https://github.com/guardstrikelab/Carla\_apollo\_bridge/issues/159.}}\footnote{\href{https://github.com/dora-rs/dora/issues/660}{https://github.com/dora-rs/dora/issues/660.} We were the first to receive official confirmation.}. Conclusively, under the DSL framework, T2S with Carla as its foundational simulation platform is compatible with supporting multiple SUT tests, thereby ensuring compatibility and support for parallel testing of a myriad of SUTs.

\begin{table}[!t]
\centering
\caption{Number of safety violation events for four different SUTs (368 scenarios, with the maximum number of violations for each type highlighted in bold).}
\label{tab:accuracy_element}
\begin{tabular}{>{\raggedright}p{0.15\linewidth}>{\raggedright\arraybackslash}p{0.2\linewidth}>{\raggedright\arraybackslash}p{0.2\linewidth}>{\raggedright\arraybackslash}p{0.2\linewidth}}
\toprule
\multicolumn{1}{c}{\textbf{ADS}} & \textbf{Rule Violation} & \textbf{Collisions} & \textbf{TimeOut} \\
\midrule
Apollo & 31 & 23 & 49 \\
Autoware & 37 & 11 & 19 \\
Interfuser & \textbf{76} & 49 & \textbf{60} \\
Dora-RS & 61 & \textbf{67} & 50 \\
\bottomrule
\end{tabular}
\end{table}

\subsection{Ablation Studies}\label{sec5.4}

To delve deeper into the influence of individual prompt engineering modules within the LLM-based parser, we conducted a series of ablation studies as depicted in Table 4. The analysis elucidates that, horizontally, there's a progressive enhancement in the matching accuracy of parsed outputs with the incremental inclusion of modules, particularly spotlighting the pivotal role of FS, which evidently steers the LLM toward generating outputs adherent to the pre-defined data structure, thereby curtailing compilation errors. For the Yi-34B-Chat, the CoT module markedly amplified its average matching accuracy by 0.159, starkly highlighting its effectiveness. Yet, the SAC module unexpectedly induced a marginal decrement, which could be speculated as stemming from the intrinsic randomness and indeterminacy associated with LLMs \cite{yin2023large}. Nonetheless, this dip was adeptly mitigated by the introduction of the SC module, which, via a voting mechanism, appreciably fortified the stability and congruence of the model's outputs. From a vertical standpoint, when subjected to rigorously crafted prompt instructions, Yi-34B-Chat registered a remarkable ascent in performance, overtaking ChatGPT-3.5 with a range expanding from a -0.025 deficit on the BP to a substantial +0.192 lead when all modules were engaged (BP-FS-CoT-SAC-SC). ChatGPT-4.0 persistently exhibited a superior intellectual performance under analogous prompt conditions, notably achieving a matching accuracy index exceeding 0.4 even in the bare BP condition. Reflecting on the performance trends, ChatGPT-4.0 exhibits a markedly slower rate of degradation in capability across various prompts compared to its counterparts, substantiating its inherently robust general intelligence. In summary, each module distinctly contributes to enhancing LLMs' output proficiency: FS systematically structures the output, CoT augments the interpretive and analytical capabilities for implicit expressions, SAC rigorously scrutinizes output for syntactic and structural integrity, and SC bolsters the overall stability and reliability of LLM-generated outputs.

\begin{table}[!t]
\centering
\caption{Ablation studies for the average matching accuracy. (Average matching accuracy = Success rate of execution $*$ Average accuracy of the element parsing.)}
\label{tab:ablation_study}
\begin{tabular}{>{\raggedright}p{0.2\linewidth}>{\raggedright\arraybackslash}p{0.12\linewidth}>{\raggedright\arraybackslash}p{0.12\linewidth}>{\raggedright\arraybackslash}p{0.12\linewidth}
>{\raggedright\arraybackslash}p{0.12\linewidth}
>{\raggedright\arraybackslash}p{0.12\linewidth}}
\toprule
\multicolumn{1}{c}{\textbf{LLM}} & \textbf{BP} & \textbf{BP-FS} & \textbf{BP-FS-CoT} & \textbf{BP-FS-CoT-SAC} & \textbf{BP-FS-CoT-SAC-SC} \\
\midrule
Yi-34B-Chat & 0.109 & 0.485 & 0.644 & 0.642 & 0.734 \\
ChatGPT-3.5 & 0.134 & 0.416 & 0.458 & 0.507 & 0.542 \\
ChatGPT-4.0 & 0.548 & 0.656 & 0.685 & 0.765 & 0.803 \\
\bottomrule
\end{tabular}
\end{table}

\subsection{Discussions and Limitations}\label{sec5.5}

T2S streamlines the construction of scenario DSL files while preserving the fidelity and precision of scenario execution, thereby signifying a tangible reduction in manual labor. Its integration with human expertise is poised to significantly enhance simulation-based testing efficiency in ADS. Nonetheless, T2S is not devoid of challenges that necessitate strategic solutions. 

\begin{itemize}
\item {{\textbf{Robustness of LLM:}} Variations in testing scenario texts, albeit semantically identical, can precipitate disparate scenario representations due to the inherent fragility of LLM robustness -- a longstanding challenge in the field of deep learning. Despite stabilization efforts through the use of SAC and SC, the resilience of LLM-based models against inconsistencies demands further refinement. Advancing the reliability of LLMs calls for additional focused training involving meticulous dataset curation and model fine-tuning.}

\item {{\textbf{Scalability Constraints:}} While the current process incentivizes the usage of pre-existing components from the element repository, hardcoding of DSL files -- though beneficial for readability, executability, and controllability -- unintentionally limits the scope for probing uncharted corner cases. Future research endeavors aspire to explore innovative strategies for adaptively generating matching scenarios in the absence of pre-established element library components. }

\item {{\textbf{Format Limitation:}} The scope of the present study confines itself to the generation of OpenScenario standard files. The specificity of the output restricts its adaptability to alternative DSL target files, thereby limiting the versatility of T2S. To overcome this bottleneck, an end-to-end scenario material library generation methodology -- which envisions training a description file writing model through exposure to a wide range of scenario descriptions -- warrants exploration. This model could independently author DSL code, shaping a new frontier in AD testing research.}
\end{itemize}

\section{Threats to Validity}\label{sec7}

We discussed threats to validity from three aspects: external validity, internal validity, and construct validity, as suggested by the literature \cite{wohlin2012experimentation}. 

\textbf{External Validity:} This pertains to the applicability and generalizability of our research methods. By employing a diverse LLM approach, we've automated the generation of test scenarios adhering to the standards of both North American NHTSA and Chinese C-NCAP, alongside expert-customized inputs. These scenarios are characterized by their cross-regional and multi-dimensional attributes and have been implemented within the simulation Carla to test ADS that follow different technological paradigms. Thanks to meticulously structured prompt engineering, we achieved an automatic scenario generation accuracy of 80.3\%, showcasing the T2S framework's adeptness in comprehending texts, matching elements, and generating scenarios. The final scenario descriptions are formatted in the internationally recognized DSL, OpenScenario, ensuring seamless adaptability to other simulation environments compatible with this standard. 

\textbf{Internal Validity:} This concerns the direct correlation between experimental outcomes and our research methodology. Through comparative ablation studies specifically on the prompt module, we observed that the combination of LLM and strategic prompt engineering significantly improves the success rate and accuracy of the compilation of autogenerated test scenarios. It reinforces the effectiveness of our approach in generating more reliable and applicable scenarios for ADS testing. 

\textbf{Construct Validity:} This relates to the suitability of the ADS and simulators employed in our study. To ensure a broad representation, we selected four ADS technologies known for their distinct operational frameworks. Autoware and Apollo are modular ADS solutions prevalent in industry applications. Interfuser, notable for its high performance in the Carla leaderboard, exemplifies a multimodal end-to-end ADS indicative of potential future trends. Dora-RS, distinct for its reliance on visual processing, surpasses conventional ROS2 communication architectures in optimization.
We use the high-fidelity simulator, Carla, as a unified platform, enabling consistent testing of the distinct ADS' safety performance to ensure construct validity.

\section{Conclusion and Future Direction}\label{sec8}

In conclusion, this research elucidates the design and implementation of an advanced automated scenario generation framework, Text2Scenario, which is meticulously engineered to process autonomous driving test simulations from natural language descriptions, utilizing an advanced large language model. The LLM-driven parser stringently selects pertinent scenario elements from our extensive hierarchical scenario repository that align with the provided textual descriptions. The integration of static and dynamic elements within a structured prioritization matrix subsequently facilitates the generation of executable DSL files. These files are instrumental in the real-time appraisal of ADS performance. Our rigorous manual evaluation process substantiates the efficacy of the T2S approach, achieving an impressive success rate of 80.3\% in generating accurate simulation scenarios across a diverse range of input texts.

This study stands at the forefront, marking an inaugural exploration into the domain of standardized DSL scenario description file generation through natural language text inputs. As we venture into the future, our focus will pivot towards the formulation of a seamless end-to-end framework for automated scenario file production. This initiative is aimed at dismantling existing barriers associated with testing material databases. Concurrently, we aspire to incorporate the synthesis of more intricate and high-risk scenarios into our research methodology. 

\section*{Declarations}
\begin{itemize}
\item \textbf{Acknowledgements} {The authors would like to appreciate the financial support of the National Key R\&D Program of China, No. 2023YFB4301802-02, the National Natural Science Foundation of China (project number: 52441202), the National Key R\&D Project of China under Grant 2022YFB4300400, the Beijing Natural Science Foundation (project number: L243008), the Ministry of Transport of PRC Key Laboratory of Transport Industry of Comprehensive Transportation Theory (project number: MTF2023002), and SHANDONG HI-SPEED GROUP CO,LTD [Grant No. HS2023B020].}

\item \textbf{Conflict of interest} {On behalf of all the authors, the corresponding author states that there is no conflict of interest.}


\end{itemize}

\newpage
\begin{appendices}

\section{Prompt Flow for Scenario Generation}\label{secA}

\small
\begin{longtable}{p{1.0\linewidth}}
\caption{Prompt flow.} \label{tab:Prompt Flow} \\
\hline
\endfirsthead

\multicolumn{1}{c}%
{{\bfseries \tablename\ \thetable{} -- continued from previous page}} \\
\hline
\endhead

\hline \multicolumn{1}{r}{{Continued on next page}} \\ 
\endfoot

\hline
\endlastfoot

\textbf{Role Setting.} Assuming you are an expert in autonomous driving testing, your task is to generate scenario representation from the following given testing scenario description text based on the Domain-Specific Language. 

\hspace*{\fill}\ 

\textbf{Hierarchical Scenario Repository}. The Hierarchical Scenario Repository provides a dictionary of scenario components corresponding to each element that you can choose from. When creating scenario representation, please first consider the following elements for each subcomponent. If there is no element that can describe a similar meaning, then create a new element yourself.

\{\textit{Dictionary of Hierarchical Scenario Repository}\}

\hspace*{\fill}\ 

\textbf{Few-Shot Examples}. Below are two examples of the input testing scenario texts and the corresponding scenario representations:

LLM Input1: \{\textit{e.g. “Unprotected left turn for traffic vehicle”}\}

LLM output1: \{\textit{e.g. Scenario Representation}\}

LLM Input2: \{\textit{e.g. Testing Scenario Text}\}

LLM output2: \{\textit{e.g. Scenario Representation}\}

Based on the above description and examples, convert the following testing scenario text into the corresponding scenario representation:
\{\textit{Testing Scenario Text}\} 

\hspace*{\fill}\ 

\textbf{Chain-of-Thought.} Let’s think step by step:

1. \textit{“left turn”} means the vehicle approaches the intersection, so we choose the common “intersection” as “Road Topology” and the “global behavior” is “turn left”;

2. \textit{“Unprotected”} means there is no specific traffic signal, such as a green arrow, to protect or ensure the safety of that turn, so we do not choose “traffic light” as “traffic sign”;

3. \textit{“Unprotected left turn”} refers to the fact that the turning vehicle does not have the right-of-way and must take extra caution to avoid a potential collision. In this case, the vehicle intending to make a left turn must yield to oncoming traffic from the opposite direction, so we choose “yield” as “longitudinal oracle”; ...

\hspace*{\fill}\ 

\textbf{Syntax Alignment Checking.}

1. Knowledge Validation: Think again if the elements in the generated output scenario representation consistent with the input testing scenario text? If not, correct the inconsistencies and output the revised scenario representation.

2. Syntax Harmonization: 

1). Check again the elements in the generated output scenario representation are from the Scenario Repository. \{\textit{Dictionary of Hierarchical Scenario Repository}\}. If you cannot find a close element in the Scenario Repository consistent with the input testing scenario text, keep your output as the answer.

2). Correct the output to dictionary data structure format.

\hspace*{\fill}\ 

\textbf{Self-consistency.}\\
\end{longtable}

In this section, we detail the prompt flow employed to instruct the Large Language Models (LLMs) for the conversion from textual description to structured scenario representation, also known as logical scenarios, as shown in Tab.\ref{tab:Prompt Flow}. We utilized the publicly accessible APIs of three advanced LLMs: Yi-34B-Chat, ChatGPT-3.5, and ChatGPT-4.0, designated by their respective versions, yi-34B-v1, gpt-3.5-turbo-0613, and gpt-4-1106-preview. Post-generation, the parameters of the DSL output were meticulously adjusted by human experts to ensure an optimal level of interaction among the various traffic participants. This manual fine-tuning process was essential to accurately capture the intricate dynamics of real-world traffic scenarios in our simulations.

\section{Prompt for Scenario Description Texts}\label{secB}

We have meticulously cataloged a comprehensive set of 92 unique scenario description texts, as delineated in Tab.\ref{tab:Scenario Description Text}. These foundational scenarios undergo systematic variations to incorporate diverse weather conditions, road topologies, and types of traffic participants. This augmentation process multiplies the base scenarios, culminating in a total of 368 distinct scenarios that encompass a broad spectrum of driving situations.

\small
\begin{longtable}{p{1.0\linewidth}}
\caption{Scenario description texts.} \label{tab:Scenario Description Text} \\
\hline
\endfirsthead

\multicolumn{1}{c}%
{{\bfseries \tablename\ \thetable{} -- continued from previous page}} \\
\hline
\endhead

\hline \multicolumn{1}{r}{{Continued on next page}} \\ 
\endfoot

\hline
\endlastfoot

\textbf{NTFSH}.\\
1. Control loss: Control loss without previous action. The ego-vehicle loses control due to bad RAINY conditions on the road and it must recover, coming back to its original lane.\\
2. Traffic negotiation: Unprotected left turn at intersection with oncoming traffic. The ego-vehicle is performing an unprotected left turn at an intersection, yielding to oncoming traffic. This scenario occurs at both signalized and non-signalized junctions.\\
3. Right turn at an intersection with crossing traffic. The ego-vehicle is performing a right turn at an intersection, yielding to crossing traffic (oncoming from the left intersection). This scenario occurs at both signalized and non-signalized junctions.\\
4. Crossing negotiation at an unsignalized intersection. The ego-vehicle needs to negotiate with other vehicles to cross an unsignalized intersection. In this situation it is assumed that the first to enter the intersection has priority. \\
5. Crossing traffic running a red light at an intersection. The ego-vehicle is going straight at an intersection but a crossing vehicle runs a red light, forcing the ego-vehicle to avoid the collision. This scenario occurs at both signalized and non-signalized junctions.\\
6. Crossing with oncoming bicycles. The ego-vehicle needs to perform a turn at an intersection yielding to bicycles crossing from either the left or right.\\
7. Highway. Highway merge from on-ramp. The ego-vehicle merges into moving highway traffic from a highway on-ramp.\\
8. Highway cut-in from on-ramp. The ego-vehicle encounters a vehicle merging into its lane from a highway on-ramp. The ego-vehicle must decelerate, brake or change lane to avoid a collision.\\
9. Static cut-in. The ego-vehicle encounters a vehicle cutting into its lane from a lane of static traffic. The ego-vehicle must decelerate, brake or change lane to avoid a collision.\\
10. Highway exit. The ego-vehicle must cross a lane of moving traffic to exit the highway at an off-ramp.\\
11. Yield to emergency vehicle. The ego-vehicle is approached by an emergency vehicle coming from behind. The ego-vehicle must maneuver to allow the emergency vehicle to pass.\\
12. Obstacle avoidance. Obstacle in lane. The ego-vehicle encounters an obstacle blocking the lane and must perform a lane change into traffic moving in the same direction to avoid it. The obstacle may be a construction site, an accident or a parked vehicle.\\
13. Door obstacle. The ego-vehicle encounters a parked vehicle opening a door into its lane and must maneuver to avoid it.\\
14. Slow moving hazard at lane edge. The ego-vehicle encounters a slow moving hazard blocking part of the lane. The ego-vehicle must brake or maneuver next to a lane of traffic moving in the same direction to avoid it.\\
15. Slow moving hazard at lane edge. The ego-vehicle encounters a slow moving hazard blocking part of the lane. The ego-vehicle must brake or maneuver to avoid it next to a lane of traffic moving in the opposite direction.\\
16. Vehicle invading lane on bend. The ego-vehicle encounters an oncoming vehicles invading its lane on a bend due to an obstacle. It must brake or maneuver to the side of the road to navigate past the oncoming traffic.\\
17. Braking and lane changing. Longitudinal control after leading vehicle’s brake. The leading vehicle decelerates suddenly due to an obstacle and the ego-vehicle must perform an emergency brake or an avoidance maneuver.\\
18. Obstacle avoidance without prior action. The ego-vehicle encounters an obstacle / unexpected entity on the road and must perform an emergency brake or an avoidance maneuver.\\
19. Pedestrian emerging from behind parked vehicle. The ego-vehicle encounters an pedestrian emerging from behind a parked vehicle and advancing into the lane. The ego-vehicle must brake or maneuver to avoid it.\\
20. Obstacle avoidance with prior action - pedestrian or bicycle. While performing a maneuver, the ego-vehicle encounters an obstacle in the road, either a pedestrian or a bicycle, and must perform an emergency brake or an avoidance maneuver.\\
21. Obstacle avoidance with prior action - vehicle. While performing a maneuver, the ego-vehicle encounters a stopped vehicle in the road and must perform an emergency brake or an avoidance maneuver.\\
22. Parking Cut-in. The ego-vehicle must slow down or brake to allow a parked vehicle exiting a parallel parking bay to cut in front.\\
23. Parking Exit. The ego-vehicle must exit a parallel parking bay into a flow of traffic. \\

\hspace*{\fill}\ 

\textbf{C-NCAP.}

1. AEB function test. The front vehicle is stationary and the ego vehicle is driving behind it to test its AEB function.\\
2. Vehicle crossing. The ego vehicle is driving normally, and a traffic car suddenly drives out of the side intersection, testing its obstacle avoidance ability.\\
3. Vehicle crossing. The ego vehicle is driving normally, and a traffic vehicle suddenly emerges from a side roadway that is obscured by a large number of vehicles, testing its obstacle avoidance ability.\\
4. Unprotected left turn. The ego vehicle wants to turn left, but the traffic car is coming from the opposite intersection.\\
5. Passing a pedestrian. The ego vehicle is driving normally, and there is a pedestrian walking ahead on the right side of the road. \\
6. Passing a two-wheeler. The ego vehicle is driving normally, and there is a bicycle in the opposite lane on the left side.\\
7. Avoiding stationary vehicles in the same lane. The ego vehicle is driving normally and discovers that there is a traffic jam ahead.  The ego vehicle changes lanes to the left to avoid it.\\
8. The ego vehicle is driving normally while several traffic vehicles are parked in the right lane.\\
9. The ego vehicle is driving normally, and there are several traffic vehicles parked on both sides of the road.\\
10. The ego vehicle is driving along a curved road, and there is a pedestrian walking on the sidewalk on the right side.\\
11. The ego vehicle is driving normally and suddenly detects a pedestrian crossing the road ahead.\\
12. The ego vehicle is preparing to turn left at the intersection, but there is a traffic car parked on the route it is preparing to drive on.
13. The ego vehicle is driving normally, and a traffic car in front suddenly turns right.\\
14. The ego vehicle is driving on the left lane of a two-lane curve, with a traffic vehicle on the right side driving at a low speed. The ego vehicle overtakes it.\\
15. The ego vehicle is preparing to change lanes to the left in the right lane of a two-lane road when suddenly a traffic vehicle drives up from the left lane.\\
16. The ego vehicle is ready to start from a pile of stopped vehicles and change lanes to the left, but there is a motorcycle coming from the left rear.\\
17. The ego vehicle is ready to start from a pile of stopped vehicles and change lanes to the left, but there is a pedestrian coming from the left rear.\\

\hspace*{\fill}\ 

\textbf{Customization from experts}.

1. An ego car was making a left turn at an intersection when a npc traffic vehicle suddenly accelerated from the right lane of ego car and changed lanes to the left, overtaking the ego car and stopping it, under very heavy rain weather conditions, with a posted speed limit of 55 mph or more.\\
2. Pedestrian crossing unexpectedly. The ego-vehicle encounters a pedestrian jaywalking or crossing the road unexpectedly, requiring immediate braking or evasive maneuvers, weather is normal cloudy.\\
3. The left side of the highway is under repair, and there is a row of cones to guide the lane change to the right with normal sunny.\\
4. Cyclist weaving. The ego-vehicle encounters a cyclist weaving between lanes or riding erratically, posing a collision risk.\\
5. Debris on the road. The ego-vehicle encounters debris or objects on the road, requiring swift lane changes or braking to avoid collision, weather is extremely foggy.\\
6. Blind intersection. The ego-vehicle approaches an intersection with limited visibility due to buildings, vegetation, or other obstructions, increasing the risk of colliding with crossing traffic.\\
7. Unprotected left turn. The ego-vehicle needs to make an unprotected left turn across oncoming traffic, requiring precise timing and gap detection.\\
8. Night-time driving. The ego-vehicle operates at night or in low-light conditions, where visibility is reduced, and hazards are harder to detect.\\
9. Adverse weather conditions. The ego-vehicle operates in heavy rain, snow, fog, or other adverse weather conditions, reducing visibility and traction.\\
10. Merging into high-speed traffic. The ego-vehicle needs to merge into a highway or high-speed traffic flow, requiring precise timing and gap detection.\\
11. Construction zone. The ego-vehicle navigates through a construction zone with temporary lane shifts, reduced lane widths, and workers or equipment near the road.\\
12. Stopped traffic ahead. The ego-vehicle encounters a sudden stop in traffic, requiring immediate braking to avoid rear-ending the vehicle in front.\\
13. Aggressive driver behavior. The ego-vehicle encounters an aggressive or erratic driver tailgating, cutting in, or making sudden lane changes.\\
14. Animal crossing. The ego-vehicle encounters an animal crossing the road, requiring immediate braking or evasive maneuvers and the weather is normal windy.\\
15. Roundabout navigation. The ego-vehicle must navigate a multi-lane roundabout, requiring precise timing and yielding to traffic already in the roundabout.\\
16. Narrow road with oncoming traffic. The ego-vehicle travels on a narrow road with limited clearance for oncoming traffic, requiring precise positioning and potential yielding.\\
17. School zone navigation. The ego-vehicle navigates through a school zone with increased pedestrian activity, lower speed limits, and potential crossing guards.\\
18. Emergency vehicle approaching. The ego-vehicle encounters an approaching emergency vehicle (police, fire, ambulance) and must yield the right-of-way.\\
19. Toll booth navigation. The ego-vehicle must navigate through a toll booth area, potentially changing lanes or stopping to pay tolls, weather is normal cloudy.\\
20. Parking lot maneuvers. The ego-vehicle must navigate through a crowded parking lot, with pedestrians, shopping carts, and vehicles backing out of spaces.\\
21. Railroad crossing. The ego-vehicle approaches an active railroad crossing and must stop for a passing train or crossing gates and the weather is normal windy.\\
22. Lane closure due to accident. The ego-vehicle encounters a lane closure due to an accident or roadwork, requiring a lane change or merging maneuver.\\
23. Pedestrian crossing at unmarked crosswalk. The ego-vehicle encounters pedestrians crossing at an unmarked crosswalk, requiring vigilance and potential yielding.\\
24. Bicyclist on road shoulder. The ego-vehicle encounters a bicyclist riding on the road shoulder, requiring increased clearance or lane change maneuvers and the weather is normal windy.\\
25. Temporary traffic control devices. The ego-vehicle encounters temporary traffic control devices (cones, barricades, flaggers) due to construction or events, requiring compliance with temporary traffic patterns.\\
26. Stopped vehicle on shoulder. The ego-vehicle encounters a stopped vehicle on the shoulder, requiring lane change maneuvers or increased clearance with little rain.\\
27. Fallen tree or power line. The ego-vehicle encounters a fallen tree or power line blocking part of the road, requiring evasive maneuvers or lane changes with heavy wind.\\
28. Intersection without traffic signals. The ego-vehicle approaches an intersection without traffic signals or stop signs, requiring careful navigation and yielding to other vehicles.\\
29. Merge from entrance ramp. The ego-vehicle must merge from an entrance ramp onto a highway or high-speed road, requiring precise timing and gap detection.\\
30. Exit from highway. The ego-vehicle must exit from a highway or high-speed road, requiring proper lane positioning and timing of the exit maneuver , weather is normal rainy.\\
31. Pedestrian crossing at mid-block crosswalk. The ego-vehicle encounters pedestrians crossing at a mid-block crosswalk, requiring vigilance and potential yielding.\\
32. Shared road with pedestrians and cyclists. The ego-vehicle travels on a road shared with pedestrians and cyclists, requiring increased awareness and caution.\\
33. Intersection with obstructed view. The ego-vehicle approaches an intersection with limited visibility due to buildings, vegetation, or parked vehicles, increasing the risk of colliding with crossing traffic.\\
34. Lane shift due to construction. The ego-vehicle encounters a temporary lane shift due to construction, requiring precise lane positioning and awareness of changing road patterns.\\
35. Pedestrian crossing at signalized intersection. The ego-vehicle approaches a signalized intersection with pedestrians crossing, requiring compliance with traffic signals and yielding to pedestrians with heavy rain.\\
36. Stopped public transit vehicle. The ego-vehicle encounters a stopped public transit vehicle (bus, tram, train) with passengers embarking or disembarking, requiring caution and potential lane changes, weather is heavy rainy.\\
37. Pedestrian crossing at unmarked mid-block location. The ego-vehicle encounters pedestrians crossing at an unmarked mid-block location, requiring increased vigilance and potential yielding and the weather is normal windy.\\
38. Merge into traffic from driveway or parking lot. The ego-vehicle must merge into traffic from a driveway or parking lot, requiring precise timing and gap detection.\\
39. Unprotected right turn on red. The ego-vehicle must make an unprotected right turn on a red light, requiring caution and yielding to pedestrians and cross traffic.\\
40. Bicyclist riding in traffic lane. The ego-vehicle encounters a bicyclist riding in the traffic lane, requiring increased clearance or lane change maneuvers.\\
41. Road debris from construction or accident. The ego-vehicle encounters debris or objects on the road from a construction site or accident, requiring evasive maneuvers or lane changes.\\
42. Intersection with obstructed view due to parked vehicles. The ego-vehicle approaches an intersection with limited visibility due to parked vehicles, increasing the risk of colliding with crossing traffic, weather is heavy rain.\\
43. Merging onto a curved highway entrance ramp. The ego-vehicle must merge onto a curved highway entrance ramp, requiring precise timing, gap detection, and lane positioning.\\
44. Pedestrian crossing at crosswalk with limited visibility. The ego-vehicle approaches a crosswalk with limited visibility due to obstructions or lighting conditions, increasing the risk of colliding with pedestrians and the weather is normal cloudy.\\
45. Motorcyclist lane splitting. The ego-vehicle encounters a motorcyclist lane splitting or filtering between lanes of traffic, requiring increased awareness and caution.\\
46. Navigating through a traffic circle or roundabout. The ego-vehicle must navigate through a multi-lane traffic circle or roundabout, requiring precise timing, lane positioning, and yielding to traffic already in the circle, weather is heavy fog.\\
47. Pedestrian crossing at intersection with obstructed view. The ego-vehicle approaches an intersection with limited visibility due to obstructions, increasing the risk of colliding with pedestrians crossing the intersection, weather is heavy fog.\\
48. Merge from parallel parking spot. The ego-vehicle must merge into traffic from a parallel parking spot, requiring precise timing, gap detection, and awareness of surrounding traffic.\\
49. Navigating through a toll plaza or toll booth area. The ego-vehicle must navigate through a toll plaza or toll booth area, requiring lane changes, potential stopping, and compliance with payment procedures, weather is heavy fog.\\
50. Pedestrian crossing at mid-block location with obstructed view. The ego-vehicle encounters pedestrians crossing at a mid-block location with limited visibility due to obstructions, increasing the risk of colliding with pedestrians.\\
51. Navigating through a parking garage or structure. The ego-vehicle must navigate through a parking garage or structure, with tight turns, potential obstacles, and pedestrians in close proximity and the weather is normal windy.\\
52. Merging onto a highway with a short acceleration lane. The ego-vehicle must merge onto a highway with a short acceleration lane, requiring precise timing, gap detection, and acceleration to match traffic speed, weather is normal wet.\\

\end{longtable}









\section{Extensive Experiment Results}\label{secC}

We have provided a more detailed number of safety violation metrics for Tab.\ref{tab:NumberSV}. Note that there may be multiple safety violations in a scenario.

\begin{itemize}
    \item {Running red lights (RRL): }{An essential aspect of traffic
law adherence.}

    \item {Running stop signs (RSS): }{Assesses how much times the vehicle fails to stop at stop signs, a key traffic rule compliance metric.}

    \item {Running speed limit (RSL): }{Assesses how much times the vehicle exceeds speed limits, a key traffic rule compliance metric.}

    \item {Lane invasion (LI): }{Quantifies the numbers of lane invasions, a measure of lane-keeping accuracy.}

    \item {Crossing solid lines (CSL): }{Quantifies the numbers of crossing solid lines, a measure of traffic rule compliance.}

    \item {Go in a wrong direction not allowed by traffic regulations (WD): }{Assesses how much times the vehicle goes in a direction not allowed by traffic regulations, a key traffic rule compliance metric.}

    \item {Violation of road rights (i.e. failure to yield to pedestrians, bicycles, emergency vehicles, etc.) (VRR): }{Measuring the cognitive ability of vehicles in terms of road rights, a more advanced indicator for evaluating driving ability.}

    \item {Out of Road (OR): }{Quantifies the times the vehicle deviates from its intended roadway, indicating lane discipline.}

    \item {Collision (C.): }{Evaluates the times of collisions, reflecting the AV’s accident avoidance capability.}

    \item {Time out (TO): }{Quantifies the times the vehicle fails to reach destinations in time, indicating driving capability.}

\end{itemize}

\begin{table}[!t]
\centering
\caption{Number of safety violations for different SUTs.}
\label{tab:NumberSV}
\begin{tabular}{>{\raggedright}p{0.15\linewidth}>{\raggedright\arraybackslash}p{0.15\linewidth}>{\raggedright\arraybackslash}p{0.15\linewidth}>{\raggedright\arraybackslash}p{0.15\linewidth}>{\raggedright\arraybackslash}p{0.15\linewidth}}
\toprule
\multicolumn{1}{c}{\textbf{Metrics}} & \textbf{Apollo} & \textbf{Autoware} & \textbf{Interfuser} & \textbf{Dora-RS} \\
\midrule
RLL &1 &0 &16 &24 \\
RSS &8 &8 &12 &12 \\
RSL &0 &2 &2 &0 \\
LI &4 &7 &10 &6 \\
CSL &4 &10 &15 &7 \\
WD &4 &4 &10 &4 \\
VRR &2 &2 &3 &2 \\
OR &8 &4 &8 &6 \\
C. &23 &11 &49 &67 \\
TO &49 &19 &60 &50 \\
\bottomrule
\end{tabular}
\end{table}

\section{Detailed Setup of Carla-X Co-Simulation}\label{secD}

We utilize Carla to facilitate traffic simulation and construct integrated simulation platforms via various bridge interfaces. This section details the methodology used to establish four SUT co-simulation platforms, collectively referred to as Carla-X.

\begin{itemize}
    \item {Apollo.} {Apollo-v8.0.0\footnote{\href{https://github.com/ApolloAuto/apollo?tab=readme-ov-file}{https://github.com/ApolloAuto/apollo?tab=readme-ov-file}} plays the role of SUT. Carla-Apollo-bridge\footnote{\href{https://github.com/guardstrikelab/carla_apollo_bridge?tab=readme-ov-file}{https://github.com/guardstrikelab/carla\_apollo\_bridge?tab=readme-ov-file}} developed by guardstrikelab is used to build communications. All parameters are configured using native settings. Apollo provides dozens of core modules, such as perception, prediction, planning, control, and human-machine interaction to achieve system level. }

    \item{Autoware.} {Autoware-universe\footnote{\href{https://github.com/autowarefoundation/autoware.universe}{https://github.com/autowarefoundation/autoware.universe}} stack is used to test. Carla-Autoware-Bridge\footnote{\href{https://github.com/guardstrikelab/carla_autoware_bridge}{https://github.com/guardstrikelab/carla\_autoware\_bridge}} developed by guardstrikelab is used to build communications. All parameters are configured using native settings. Autoware also adopts a modular design ethos similar to Apollo. }

    \item{Interfuser.} {Interfuser\footnote{\href{https://github.com/opendilab/InterFuser}{https://github.com/opendilab/InterFuser}} uses multimodal fusion perception data output to plan routes. We directly use the pretrained weights provided by the author. The model was trained on 7 maps and 10 weather conditions. The epoch is 25, the warm-up epoch is 5, the learning rate is 0.0005, the batch size is 16, the weight decay is 0.05, the backbone learning rate is 0.0002, and the input size of multiview is $3\times128\times128$. }

    \item {Dora-RS.} {Dora-RS\footnote{\href{https://github.com/dora-rs/dora}{https://github.com/dora-rs/dora}} is a fast and simple dataflow-oriented robotic framework with perception and control capabilities. The Yolo-v5\footnote{\href{https://github.com/ultralytics/yolov5}{https://github.com/ultralytics/yolov5}}, Frenet Optimal Trajectory\footnote{\href{https://github.com/erdos-project/frenet_optimal_trajectory_planner}{https://github.com/erdos-project/frenet\_optimal\_trajectory\_planner}} and PID\footnote{\href{https://github.com/Dlloydev/QuickPID}{https://github.com/Dlloydev/QuickPID}} represent the fundamental algorithms underpinning perception, planning and control. }
    
\end{itemize}





\end{appendices}





\end{document}